\documentclass[twocolumn]{aastex62}

\usepackage{amsmath}

\hypersetup{linkcolor=black,citecolor=black,filecolor=cyan,urlcolor=magenta}

\def\plus/{\texttt{+}}
\def\ro/{$\rho$ Ophiuchi}
\def\roa/{$\rho$ Oph A}
\def\rdc/{$\mathcal{R}_{DC}$}
\def\rdcm/{$\mathcal{R}_{DC,m}$}

\graphicspath{{./}{figures/}}

\providecommand{\sorthelp}[1]{}

\submitjournal{ApJ}
\accepted{July 18, 2019}

\shorttitle{\ro/ observations with HAWC\plus//SOFIA}
\shortauthors{Santos et al.}

\begin{document}

\title{The far-infrared polarization spectrum of $\rho$ Ophiuchi A from HAWC\plus//SOFIA observations}




\correspondingauthor{Fabio P. Santos}
\email{fabiops@mpia.edu}

\author[0000-0002-9650-3619]{Fabio P. Santos}
\affil{Center for Interdisciplinary Exploration and Research in Astrophysics (CIERA), and Department of Physics \& Astronomy, Northwestern University, 2145 Sheridan Rd, Evanston, IL, 60208, USA}
\affil{Max-Planck-Institute for Astronomy, K\"onigstuhl 17, 69117 Heidelberg, Germany}

\author[0000-0003-0016-0533]{David T. Chuss}
\affil{Department of Physics, Villanova University, 800 E. Lancaster Ave., Villanova, PA 19085, USA}

\author{C. Darren Dowell}
\affil{NASA Jet Propulsion Laboratory, California Institute of Technology, 4800 Oak Grove Drive, Pasadena, CA 91109, USA}

\author[0000-0003-4420-8674]{Martin Houde}
\affil{Department of Physics and Astronomy, University of Western Ontario, 1151 Richmond Street, London, ON N6A 3K7, Canada}

\author[0000-0002-4540-6587]{Leslie W. Looney}
\affil{Department of Astronomy, University of Illinois, 1002 West Green Street, Urbana, IL 61801, USA}

\author{Enrique Lopez Rodriguez}
\affil{SOFIA Science Center/Universities Space Research Association}
\affil{NASA Ames Research Center, M.S. N232-12, Moffett Field, CA, 94035, USA}

\author{Giles Novak}
\affil{Center for Interdisciplinary Exploration and Research in Astrophysics (CIERA), and Department of Physics \& Astronomy, Northwestern University, 2145 Sheridan Rd, Evanston, IL, 60208, USA}

\author{Derek Ward-Thompson}
\affil{Jeremiah Horrocks Institute, University of Central Lancashire, Preston PR1 2HE, United Kingdom}

\author{Marc Berthoud}
\affil{Engineering + Technical Support Group, University of Chicago, Chicago, IL 60637, USA}
\affil{Center for Interdisciplinary Exploration and Research in Astrophysics (CIERA), and Department of Physics \& Astronomy, Northwestern University, 2145 Sheridan Rd, Evanston, IL, 60208, USA}

\author{Daniel A. Dale}
\affil{Department of Physics \& Astronomy, University of Wyoming, Laramie, WY, USA}

\author[0000-0001-8819-9648]{Jordan A. Guerra}
\affil{Department of Physics, Villanova University, 800 E. Lancaster Ave., Villanova, PA 19085, USA}

\author[0000-0001-6350-2209]{Ryan T. Hamilton}
\affil{Lowell Observatory, 1400 W Mars Hill Rd, Flagstaff, AZ 86001, USA}

\author[0000-0002-8702-6291]{Shaul Hanany}
\affil{School of Physics and Astronomy, University of Minnesota / Twin Cities, Minneapolis, MN, 55455, USA}

\author{Doyal A. Harper}
\affil{Department of Astronomy and Astrophysics, University of Chicago, Chicago, IL 60637, USA}

\author{Thomas K. Henning}
\affil{Max-Planck-Institute for Astronomy, K\"onigstuhl 17, 69117 Heidelberg, Germany}

\author{Terry Jay Jones}
\affil{Minnesota Institute for Astrophysics, University of Minnesota, 116 Church St. SE, Minneapolis, MN 55455, USA}

\author{Alex Lazarian}
\affil{Department of Astronomy, University of Wisconsin, Madison, WI 53706, USA}

\author[0000-0003-3503-3446]{Joseph M. Michail}
\affil{Department of Astrophysics and Planetary Science, Villanova University, 800 E. Lancaster Ave., Villanova, PA 19085, USA}
\affil{Department of Physics, Villanova University, 800 E. Lancaster Ave., Villanova, PA 19085, USA}

\author[0000-0002-6753-2066]{Mark R. Morris}
\affil{Department of Physics and Astronomy, University of California, Los Angeles, Box 951547, Los Angeles, CA 90095-1547 USA}

\author{Johannes Staguhn}
\affil{Dept. of Physics \& Astronomy, Johns Hopkins University, Baltimore, MD 21218, USA}
\affil{NASA Goddard Space Flight Center, Greenbelt, MD 20771, USA}

\author{Ian W. Stephens}
\affil{Harvard-Smithsonian Center for Astrophysics, 60 Garden Street, Cambridge, MA, USA}

\author[0000-0002-8831-2038]{Konstantinos Tassis}
\affil{Department of Physics and ITCP, University of Crete, Voutes, GR-70013 Heraklion, Greece}
\affil{IESL and Institute of Astrophysics, Foundation for Research and Technology-Hellas, 100 N. Plastira, Voutes, GR-70013 Heraklion, Greece}

\author{Christopher Q. Trinh}
\affil{USRA/SOFIA, NASA Armstrong Flight Research Center, Building 703, Palmdale, CA 93550, USA}

\author{Eric Van Camp}
\affil{Center for Interdisciplinary Exploration and Research in Astrophysics (CIERA), and Department of Physics \& Astronomy, Northwestern University, 2145 Sheridan Rd, Evanston, IL, 60208, USA}

\author{C. G. Volpert}
\affil{University of Chicago, Chicago, IL 60637, USA}
\affil{University of Maryland, College Park, MD 20740, USA}

\author[0000-0002-7567-4451]{Edward J. Wollack}
\affil{NASA Goddard Space Flight Center, Greenbelt, MD 20771, USA}

\begin{abstract}

We report on polarimetric maps made with HAWC\plus//SOFIA toward \roa/, the densest 
portion of the \ro/ molecular complex. 
We employed HAWC\plus/ bands C ($89\,\mu$m) and D ($154\,\mu$m). 
The slope of the polarization spectrum was investigated 
by defining the quantity 
$\mathcal{R}_{DC} = p_{D}/p_{C}$, where $p_{C}$ and $p_{D}$ represent polarization degrees
in bands C and D, respectively. 
We find a clear correlation between \rdc/ and the molecular hydrogen column density 
across the cloud. A positive slope (\rdc/ $>$ 1) dominates the lower density and well 
illuminated portions of the cloud, that are heated by the high mass star Oph S1, 
whereas a transition to a negative slope (\rdc/ $<$ 1) is observed toward the denser and less evenly illuminated 
cloud core. We interpret the trends as due to a combination of: 
(1) Warm grains at the cloud outskirts, which are efficiently aligned 
by the abundant exposure to radiation from Oph S1, as proposed in the
radiative torques theory; and (2) Cold grains deep in the cloud core, which 
are poorly aligned due to shielding from external radiation.  
To assess this interpretation, we developed a very simple toy model using a spherically symmetric 
cloud core based on {\it Herschel} data, and verified that the predicted variation 
of \rdc/ is consistent with the observations.
This result introduces a new method that can be used to probe the grain alignment efficiency
in molecular clouds, based on the analysis of trends in the far-infrared polarization spectrum.


\end{abstract}



\section{Introduction}
\label{s:introduction}

It is generally believed that the magnetic field that permeates the interstellar medium (ISM) play 
an important role in the formation of stars and 
planets \citep[e.g.,][]{2006mou,mckee2007,2014krumholz,2014li}. 
The field has a tendency to become ``frozen into'' the partially ionized interstellar gas. 
In regions where the magnetic field energy density is subdominant to the kinetic energy 
of the matter, gas dynamics will influence magnetic field morphology.
This effect has been observed for H{\sc i} shells
\citep{heiles1998,fosalba2002,santos2011,frisch2017,2018soler} as well as at the edges 
of H{\sc ii} regions \citep{pavel2012,santos2012,santos2014,2016planckxxxiv}.  Another 
crucial effect is the force exerted by the field on the gas. This force may generate filamentary 
structure by guiding of turbulent gas flows \citep{1998nagai,2008nakamura,2015li}, and it 
may suppress the fragmentation of filaments, thereby partly explaining the low efficiency of 
the star formation process \citep{2004hosking,2013myers}.  
Observations of interstellar polarization are the most widely adopted technique to map the magnetic
field in the ISM, but in order to use dust polarimetry to study magnetic fields, one needs to 
understand magnetic alignment of interstellar dust grains.
In recent years it has become feasible to carry out detailed, large-scale observations 
of the plane-of-sky orientations of the magnetic field permeating the relatively denser, 
molecular phases of the ISM, by exploiting grain alignment.
Dust alignment can be detected via polarimetry of starlight that has been transmitted through a 
medium of aligned grains \citep[e.g.,][]{hall1949,hiltner1949,serkowski1975,heiles2000} or, 
more directly, by observing the polarized emission from the grains themselves 
\citep[e.g.,][]{2000hildebrand,2007page,soler2016,2016fissel,2019chuss}.

The mechanism believed to be responsible for grain alignment is referred to as Radiative 
Torques, also known as B-RATs or simply RATs mechanism
(for example, see \citealt{1976dolginov,draine_wein_1996,draine_wein_1997,2007lazarianhoang,2008hoang}, and the reviews by \citealt{lazarian2007} and \citealt{andersson2015}). 
RATs theory 
posits individual non-spherical grains spinning about their short axes and having non-zero helicity. 
When surrounded by an anisotropic field of radiation having wavelength comparable to the grain size, 
such grains are expected to spin up, precess around the magnetic field, and gradually align with 
their shorter axes preferentially parallel to the magnetic field direction. From a purely observational 
perspective, the ease with which grains become aligned with the magnetic field is seen to be influenced by 
their size \citep{kim1995}, possibly by their composition \citep{smith2000,chiar2006,2018lazarian}, and probably by 
their radiative environment, as discussed in detail below.  Observational constraints such as these 
tend to be consistent with RATs theory but many open questions remain \citep[e.g.,][]{2010andersson,andersson2011,andersson2015,2018ashton}.

For molecular sight-lines, we observe an anti-correlation between the polarization fraction of dust 
emission (or equivalently, the polarization fraction per unit optical depth for the case of polarization 
by selective extinction) and the column density \citep[e.g.,][]{1998arce,whittet2008,2016fissel,2017santos}.
Possible explanations for this observed anti-correlation are: (1) a greater degree of field disorder 
within the observed beam
along high-column density sight-lines; (2) a loss of polarizing efficiency for grains deep in molecular 
clouds; (3) a combination of both effects. The loss of polarization efficiency explanation 
could result from either changes in intrinsic grain properties for dense
molecular regions (e.g., larger or rounder grains deep in clouds) or inefficient grain alignment for
regions that are well-shielded from radiation.  Indeed, RAT theory would predict that a loss of 
grain alignment should occur for locations that are so well shielded that not even near-infrared light from 
the interstellar radiation field can penetrate \citep{alves2014,jones2015,andersson2015}.  
In addition, near-infrared spectro-polarimetry of the Taurus molecular cloud by \citet{whittet2008} shows that along with 
loss of polarization fraction, the denser sight-lines exhibit a change in the wavelength dependence of the 
polarization that is consistent with a reduction in the fraction of grains that are aligned.  They 
argue in favor of reduced grain alignment for well shielded regions as the main culprit for the 
anti-correlation, rather than field disorder or grain shape. 

In this paper, we focus on a relatively unexplored observable, which is the polarization spectrum of the 
grains' emission.  In other words, we are concerned here with the fractional polarization as a function of 
wavelength.  For the coldest regions of star forming molecular clouds, dust temperatures are in the 
range of $\sim10-15\,$K, so the dust thermal radiation is mainly in the submillimeter and 
millimeter (peaking at $\approx 300\,\mu$m).  
However, dust temperatures can be much higher near newly formed early-type stars, and from these hotter 
dust grains we expect copious far-IR radiation at relatively shorter wavelengths ($\sim 50 - 200\,\mu$m).  Using polarimetry from the Kuiper Airborne Observatory, \citet{1999hildebrand} measured 
the first far-IR polarization spectra and found that 
the polarization fraction decreases with wavelength.  After showing that the theoretically expected spectra 
for the simplest dust models were basically flat, they presented an idea for how to produce falling 
spectra: imagine that we have two kinds of regions along the same line of sight, 
namely some cold regions far from newly formed stars and hot regions closer to such sources.  
Further assume that the dust grains in the hot regions are much better aligned than the colder grains, 
perhaps due to RATs from these same sources. 
Since the hot regions with well aligned grains will be relatively brighter 
than the cold regions at the 
shorter wavelengths, we expect shorter wavelengths to be more polarized.
Therefore, in this case it is clear that the polarization 
fraction will fall with increasing wavelength, i.e., we will find negatively sloped polarization 
spectra, as observed by \citet{1999hildebrand}. 

During the two decades that have elapsed since the work by \citet{1999hildebrand}, much observational 
work has been done on far-infrared and submillimeter polarization spectra of star forming clouds, 
but the situation 
here is somewhat muddled.  Ground-based observations have found positively sloped spectra in the submillimeter, 
suggesting a minimum in the polarization spectrum near $350\,\mu$m \citep{2002vai,2008vai, 2012vai, 2013zeng}, 
but balloon-borne observations by the BLAST collaboration found flat submillimeter polarization spectra
\citep{2016gandilo,2019shariff}.  The difference may be due to the different column density regimes 
studied \citep{2016gandilo}.  There has also been progress on the theoretical side.  \citet{bethell2007} 
predicted that molecular clouds should have positive-slope polarization spectra in the 
far-IR ($\lambda < 350\,\mu$m) and flat 
spectra in the submillimeter ($\lambda > 350\,\mu$m).  \citet{draine2009} modeled the diffuse ISM, 
finding a similar situation. 
\citet{2018guillet} also modeled the diffuse ISM, finding that they could match the flat 
submillimeter-millimeter polarization spectra recently observed for the relatively tenuous regions by 
the Planck satellite \citep{planckXXII} and by BLAST \citep{2018ashton}.  
From the theoretical perspective, the positive slopes in the far-IR are attributed to: 
(1) the fact that larger-sized grains ($\gtrapprox 0.2\,\mu$m) are relatively more efficiently 
aligned as compared to 
smaller-sized grains ($\lessapprox 0.2\,\mu$m) -- this has been observationally verified by \citet{kim1995} and is 
also predicted from the RATs theory \citep[e.g., ][]{lazarian2007}; and (2) the fact that
different grain size populations follow different temperature distributions (even when subject to 
a uniform radiation field), with smaller grains being relatively warmer than larger grains due to 
their inefficiency in cooling radiatively \citep[][]{1999li_goldsmith}. As a result, 
the shorter wavelength emission 
within the far-infrared spectral range is dominated by warmer and relatively poorly aligned
small grains (i.e., less polarized), while at long wavelengths, the emission from larger and 
better aligned grains is more significant (i.e., more polarized). In addition, 
composition can also play an important role \citep[e.g.,][]{draine2009}.


The newly commissioned HAWC\plus/ far-IR polarimeter for SOFIA \citep{2010dowell,Harper2018} 
allows us to revisit 
the topic of molecular cloud far-IR polarization spectra, but with better angular resolution and 
sensitivity.  In this work, we present polarimetric observations of the nearby star forming region 
\roa/ obtained with HAWC\plus/ at two different far-IR wavelengths, $89\,\mu$m and $154\,\mu$m.  
The \roa/ region is part of L1688, which in turn is part of the Ophiuchus molecular cloud.  The 
distance to L1688 has been measured to be $137 \pm 1\,$pc by \citet{2017ortiz} via radio 
parallax measurements of 12 young stellar systems associated with this cloud. 
\roa/ exhibits wide ranges of both temperature and column density \citep{1998motte}, providing 
the opportunity to search for systematic variations in the polarization spectrum slope as a function 
of these parameters. The V-band extinction can reach levels larger than $100\,$mag 
\citep{2017friesen}.  Located at just $1.64\,$arcmin from the peak density (projected $0.065\,$pc),
the high-mass star Oph S1 warms up the surrounding environment, causing a large temperature 
gradient (from approximately $20\,$K at the core to around $40\,$K near Oph S1).  Oph S1 is the 
main heat source for \roa/, and is also associated with a $20\arcsec$ ultra-compact H{\sc ii} region
\citep{1988andre}.  The star is of type B3/4.  It has almost reached the main sequence, and 
is now transitioning from a Herbig AeBe star into a magnetic B star \citep{1988andre,2003hamaguchi}. 
VLBA parallax observations show that Oph S1 lies at a distance of $138 \pm 2\,$pc \citep{2017ortiz}.  
Taking measurement errors into account, this is consistent with the distance measured by the
same authors to L1688 as a whole (see above). \roa/ contains numerous young stellar objects 
at different evolutionary stages (Classes 0, I, II, and III) as well as a population of starless 
cores \citep{1998motte,2015pattle,2015liseau}.

In Section \ref{s:obs} below, we describe the data acquisition and reduction, and we detail the 
selection criteria used to identify data suitable for analysis.  In Section \ref{s:resan} we describe 
this analysis, after first presenting our total intensity maps and polarization measurements.  We 
discuss the results of our analysis in Section \ref{s:discussion}.  Finally, Section \ref{s:conclusions} 
is a summary.

   \begin{figure*}[!t]
   \centering
   \includegraphics[width=0.49\textwidth]{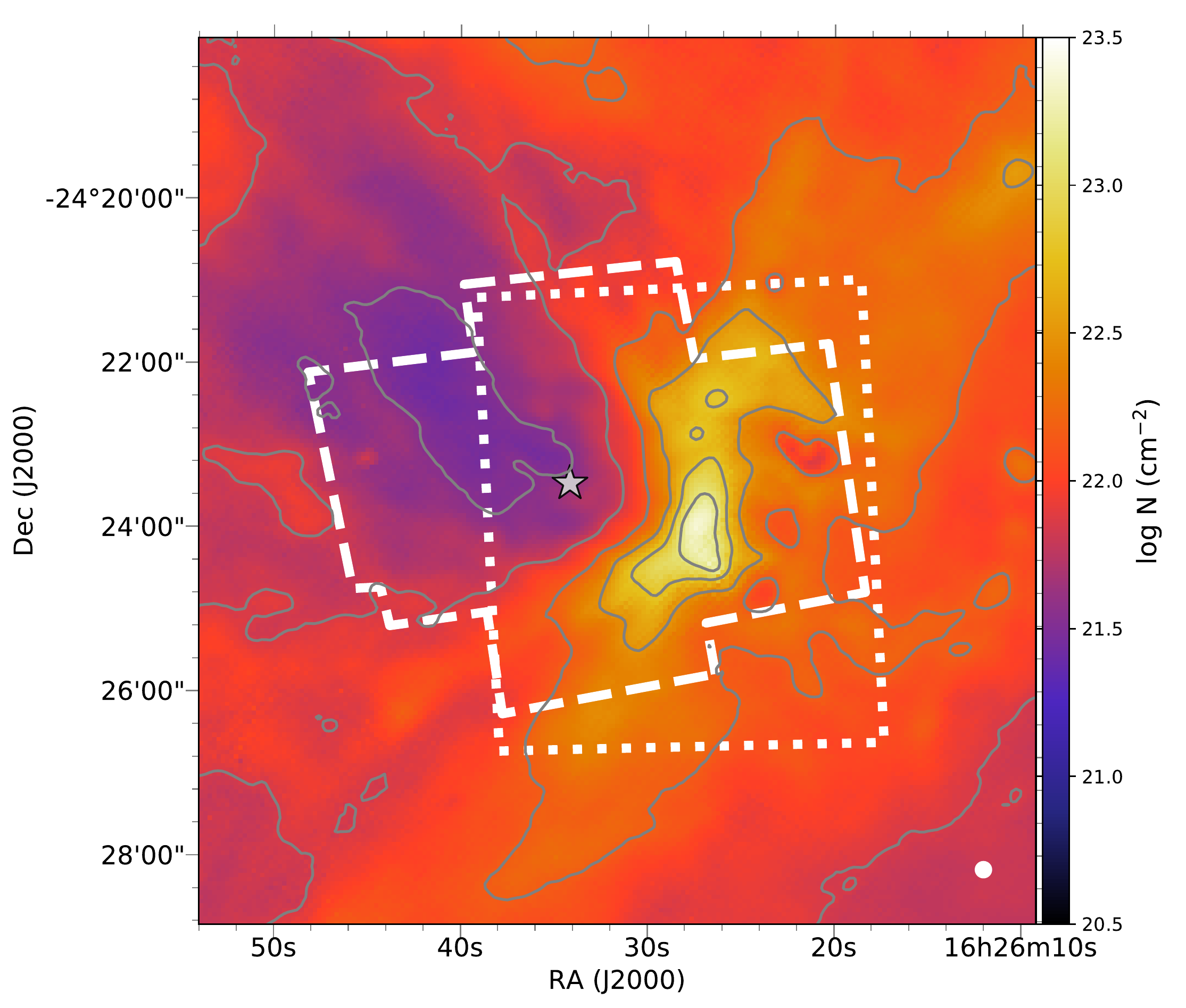} 
   \includegraphics[width=0.49\textwidth]{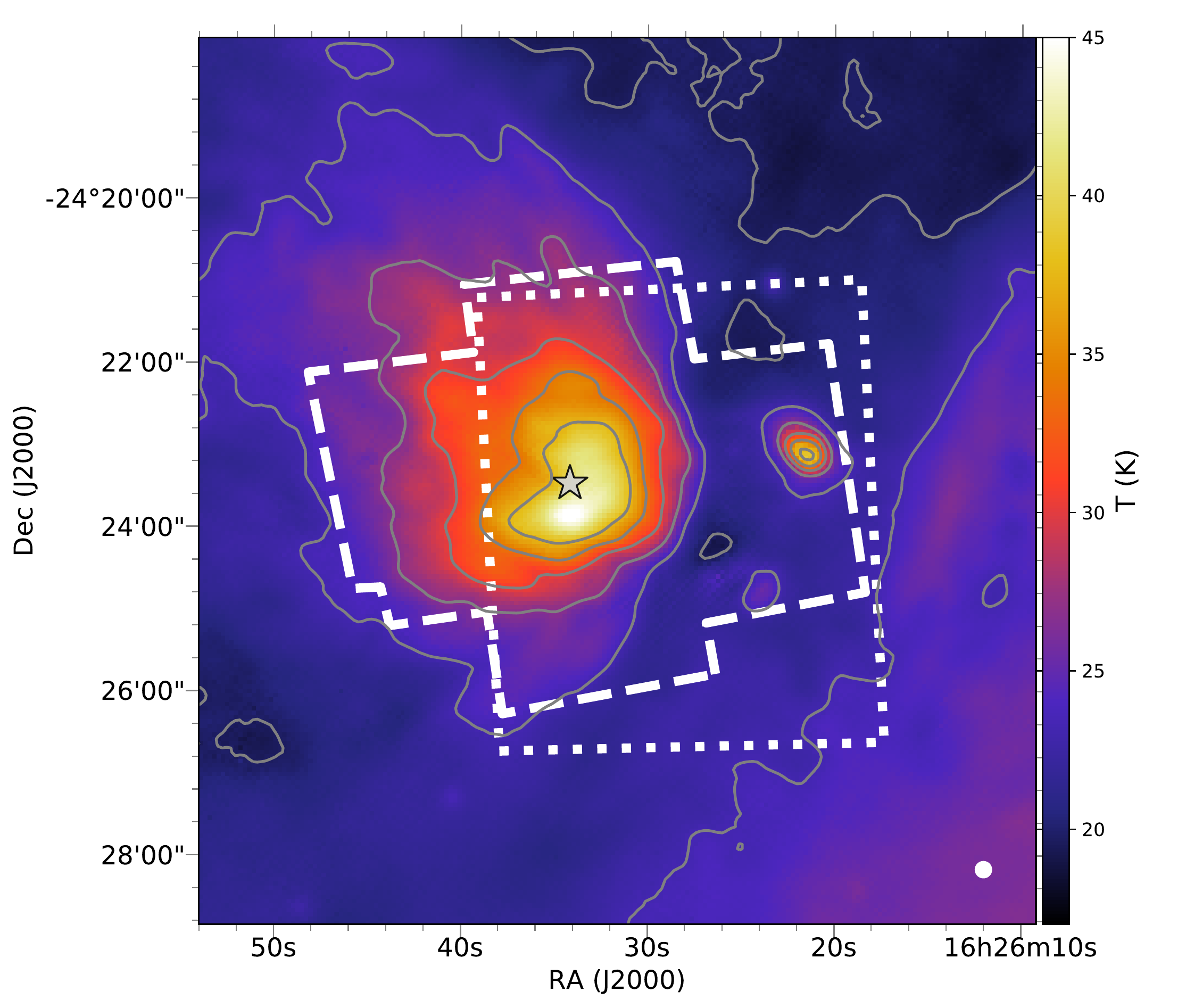} 
   \caption{\roa/ maps of the inferred molecular hydrogen (H$_{2}$) column density ($N$, left), 
   and dust temperature ($T$, right), as derived from {\it Herschel} fluxes 
   (see Section \ref{s:compdata}). The gray-colored star symbol indicates the position
   of Oph S1. The white dashed line and the white dotted line indicate 
   the areas covered by HAWC\plus/ bands C and D observations, respectively.
   In the left panel, the top contour represents $\log N(\mathrm{cm}^{-2}) = 23.1$,
   and the lower contours follow in subsequent steps of $\log N(\mathrm{cm}^{-2}) = 0.31$.
   In the right panel, the top contour represents a level of $T = 37.5\,$K, with lower 
   contours following in steps of $T = 3.6\,$K. The circle at the bottom right 
   of each panel represents the beam size (FWHM) of the {\it Herschel} data used to construct
   these maps.
   }
   \label{f:ntmaps}
   \end{figure*}

\section{Observations}
\label{s:obs}

Polarimetric data for \roa/ were obtained using HAWC\plus//SOFIA \citep{Harper2018} 
on May 5 2017 as part of the Guaranteed Time Observing 
(GTO) program. We used HAWC\plus/ bands C ($89\,\mu$m) and D ($154\,\mu$m)
that nominally provide angular resolutions of $7.8\arcsec$ and $13.6\arcsec$ FWHM (full width at half maximum), respectively.
The standard matched-chop-nod method was used \citep{2000hildebrand} with a chopping frequency of $10.2$ Hz.  For both bands, we used a chop angle 
of 154$^\circ$ (measured from equatorial North and increasing to the East) and a chop throw 
of $480\arcsec$.  These values were chosen in order to minimize the flux levels in our reference beams. A typical polarimetric observing block with HAWC\plus/ consists of a set of 4 dithered observations, 
i.e., 4 independent pointings slightly displaced from each other forming a square pattern on the 
plane of the sky. Each such block is completed in approximately 15 minutes.
The dithering displacement between individual observations was $24\arcsec$ for band C and $40\arcsec$
for band D. 

The sky areas covered by the HAWC\plus/ observations are indicated in Figure \ref{f:ntmaps} using 
dashed lines (band C) and dotted lines (band D).  This figure also shows, via color
maps, the distributions of molecular hydrogen column density ($N$, left) and dust temperature ($T$, right) as derived from archival {\it Herschel} Space Observatory dust emission data \citep{2010pilbratt}. 
Details of the methods we used to derive these maps are given in Section \ref{s:compdata}.  For band C we obtained four observing blocks, with each observing block centered on a different sky position in order to increase total sky coverage.  There was significant spatial overlap between the blocks.  The sky area covered by our band C observations (dashed lines in Figure \ref{f:ntmaps}) is approximately 
centered on Oph S1, which roughly coincides with the location of peak dust temperature.  For band D, 
two observing blocks were used, both with the same pointing. The band D map (dotted lines in Figure \ref{f:ntmaps}) is approximately centered on the cloud's column density peak.

The polarimetry data presented here were processed using the HAWC\plus/ data reduction 
pipeline version \textsc{v1.3.0-beta3} (April 2018).  As summarized by \citet{Harper2018}, the pipeline consists of a series of sequential data processing steps, 

which we briefly describe in the list below: 
\begin{itemize}
\setlength\itemsep{0.0em}
\item Demodulation of the chopped data and removal 
of bad samples (due to tracking issues, half-wave-plate and nod movements, etc.); 
\item Flat-fielding 
of the demodulated data 
based on scans across extended sources, which were then transferred to 
our observations via flux measurements of an internal calibrator;
\item Computation of the difference and sum of 
signals reflected and transmitted by the polarizer; 
\item Combination of fluxes from different nod positions and calculation of 
images for each of the Stokes parameters $I$, $Q$, and $U$; 
\item Application of astrometric corrections based on known pointing offsets during the 
observations; 
\item Correction for instrumental polarization; 
\item Rotation of the Stokes $Q$ and $U$ matrices from the instrumental to the equatorial frame; 
\item Flux correction based on a standard atmospheric opacity model; 
\item Flux calibration using observations of planets during the same flight series;
\item Application of small flux offsets to individual observations 
in order to equalize their relative background flux levels;
\item Combination of individual observations into final $I$, $Q$, and $U$ maps using 
a re-gridding and Gaussian smoothing technique \citep{2007houde}; 
\item Computation of final polarization degree 
\citep[including polarization de-biasing, e.g.,][]{wardle1974} and angle maps; 
\item Construction of polarization maps with vectors overlaid to the Stokes $I$ image. 
\end{itemize}

To check for internal consistency between different observing blocks, 
we calculated $\chi^2$ maps for $I$, $Q$, and $U$.  The result showed that 
nominal uncertainties computed from the sample variance 
underestimate the scatter in the measurements 
by about $38\%$, so we inflated the errors accordingly.  We rejected
measurements that failed the $p/\sigma_p>3$ criterion, where $p$ is the polarization degree and $\sigma_p$ is the associated uncertainty. In addition, we rejected portions 
of the maps possibly affected by reference beam contamination. The 
two reference areas are symmetrically located at the two sides of the 
central observed region, at angular distances of $480\arcsec$ from it.
\roa/ contains significant extended emission, which means that diffuse areas 
of the map may be too contaminated by the flux from the reference regions,
and must be rejected. In order to do this we estimated the band C and D flux maps for 
the two reference regions $I_1$ and $I_2$, based on modified black-body fits 
using {\it Herschel} data. The SED fitting procedure is identical to the 
one described below in Section \ref{s:compdata}. We then computed average 
reference region maps for bands C and D, 
$I_{\mathrm{ref},\lambda} = (I_{1} + I_{2})/2$. We require the main source
to have a total flux at least ten times larger than the average fluxes from the 
reference regions, i.e., we rejected measurements with 
$I_{\lambda} < 10I_{\mathrm{ref},\lambda}$, where $I_{\lambda}$ are the 
calibrated Stokes $I$ observations from HAWC\plus/. 
This cut removes detections from low flux areas near the map borders which
were showing high polarization values (typically larger than $15\%$).
The final maps contain 1717  and 906 detections of polarization for bands C and D, respectively, for Nyquist spatial sampling of polarization measurements. 

\section{Results and Analysis}
\label{s:resan}

\subsection{Overview of HAWC\plus/ intensity and polarization maps}
\label{s:maps}

   \begin{figure*}[!t]
   \centering
   \includegraphics[width=0.70\textwidth]{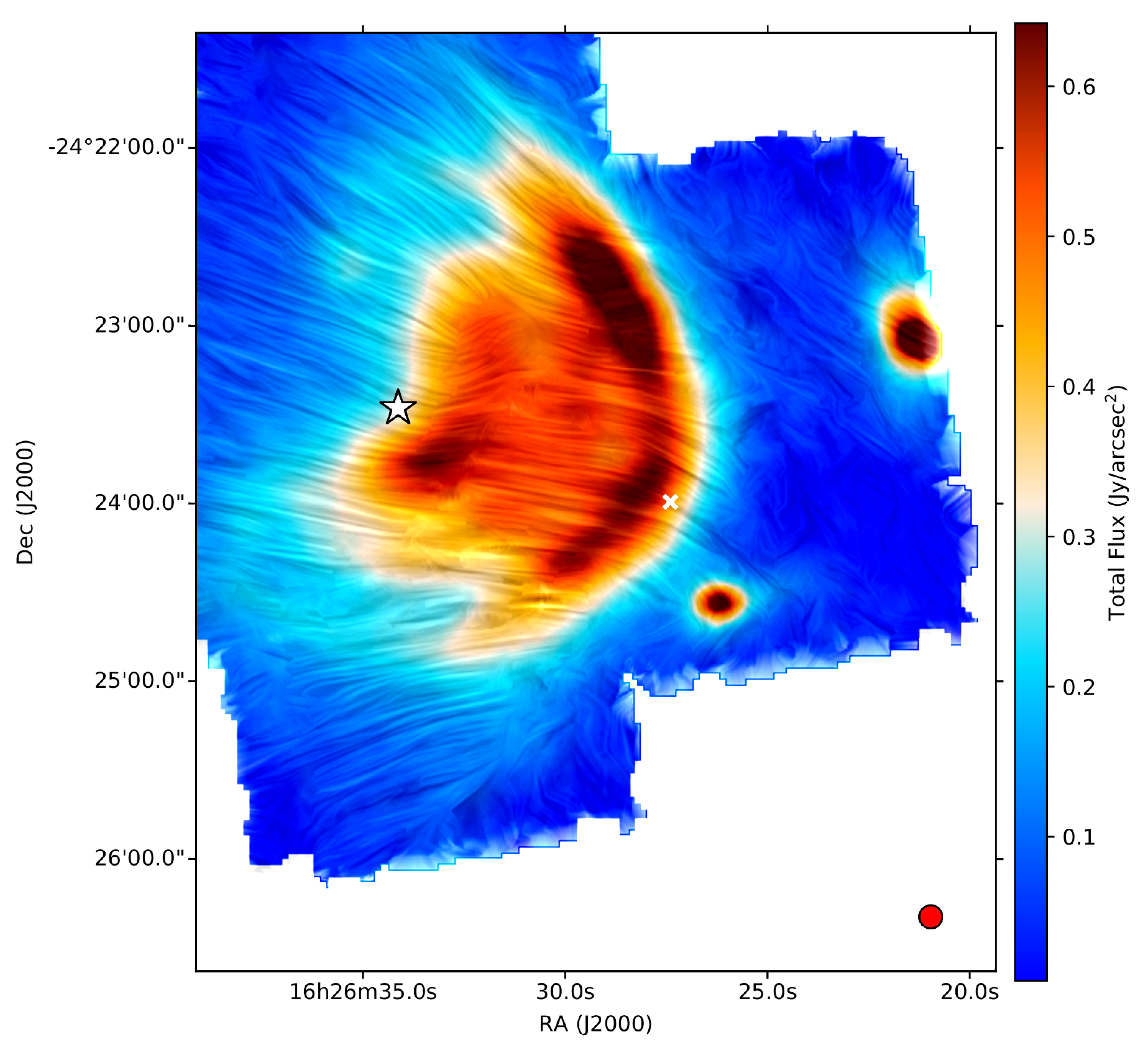} \\
   \includegraphics[width=0.70\textwidth]{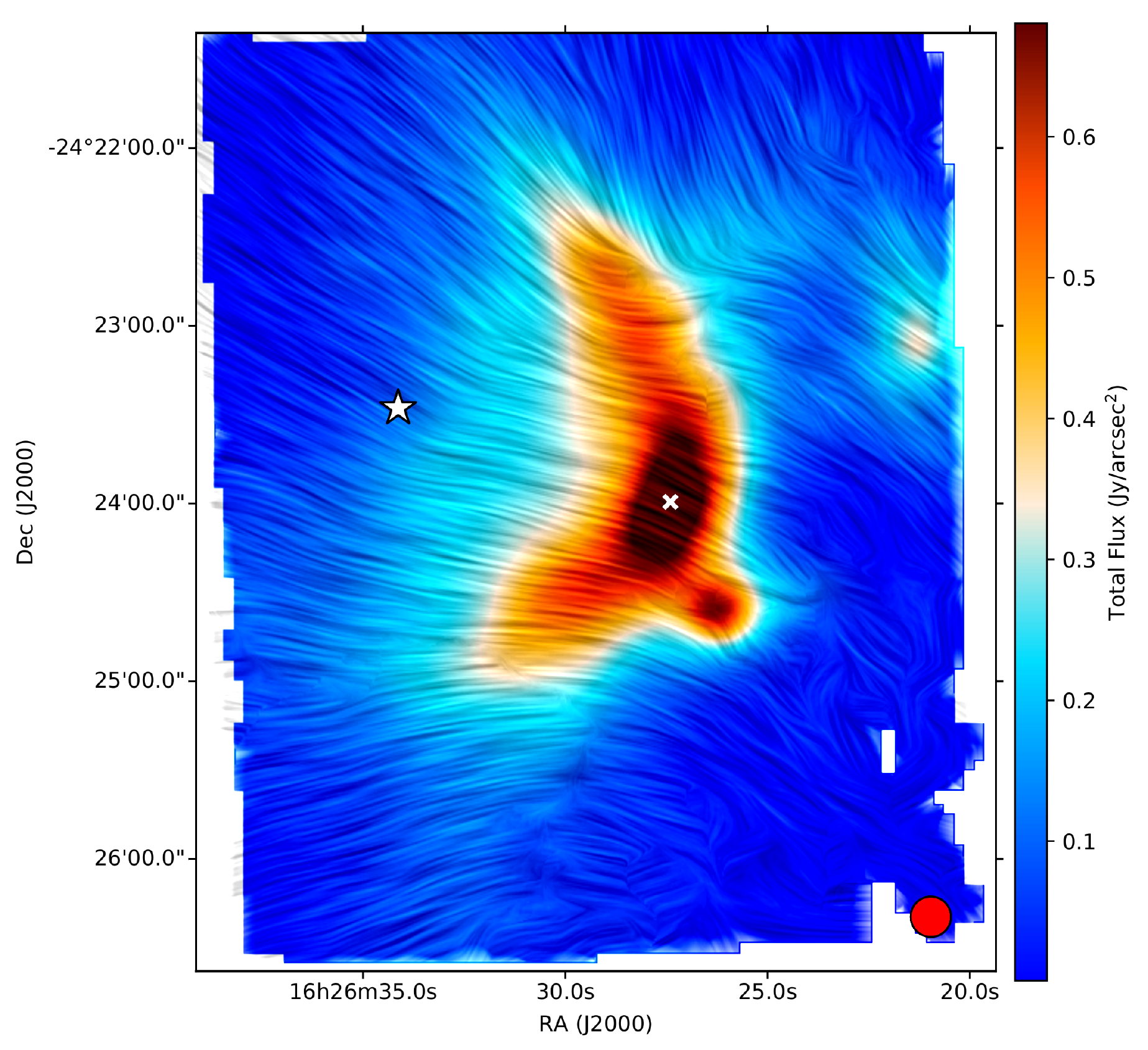} 
   \caption{Line-integral-convolution maps of the inferred magnetic field 
   direction in \roa/ from bands C (top) and D (bottom). The colors indicate
   the Stokes I emission in each band. The 
   star indicates the position of Oph S1, and the $\times$
   marks the peak column density position. Notice that both maps 
   have the same spatial scale and centering.
   The circle at the bottom right 
   of each panel represents the beam size (FWHM). 
              }
         \label{f:lic}
   \end{figure*}

From the polarization angle maps, we have generated maps of magnetic field 
direction using the standard assumption that the E vector of far-IR polarization 
is perpendicular to the component of the magnetic field in the plane of the 
sky \citep[e.g.,][]{lazarian2007}.  The maps of magnetic field direction are visualized 
using the Line-Integral-Convolution technique \citep[LIC,][]{cabral1993} in 
Figure \ref{f:lic} for bands C (top) and D (bottom).
In each map, colors represent total intensity (Stokes I)
for the respective band, while the   
overlaid LIC ``texture'' represents the inferred sky-projected 
magnetic field directions obtained from the 90-deg rotated polarization angles. 

The total intensity maps for both bands exhibit 
large arc-shaped  
features approximately
centered on Oph S1.
However, a comparison of these two total intensity maps reveals that they have important differences.
Presumably, these differences are explained by strong temperature
gradients that arise due to the effects of radiation from Oph S1. 
Band D generally has higher fluxes near the column 
density peak, preferentially probing the largest column densities. 
In contrast, the highest fluxes in band C do not correspond to the 
column density peak, but instead to the warmer areas near Oph S1. 

From the LIC maps, we note that, overall, the projected magnetic field is 
approximately perpendicular 
to the ridge observed in the column density map of Figure \ref{f:ntmaps}.
This is in qualitative 
agreement with ground-based polarimetry at longer wavelengths, as can be seen
by comparison with Figure 29 of \citet{dotson2010} and Figure 5 of \citet{kwon2018}. 
There is also a general tendency for the magnetic field in the lowest density 
gas to extend from the Southwest to Northeast direction, as was seen by 
\citet{2015kwon} in the near-infrared. However, our data, which probe deeper 
into the higher density material, see more of a curvature to the field in the 
immediate vicinity of \roa/ itself, with field lines bending perpendicularly 
to the curved ridge. This effect is seen more markedly in band D, which traces 
more of the colder material, deeper into the ridge, as compared to band C, which 
provides a better probe of the warmer material near Oph S1 and in the dense ridge 
outer layers. This gives us some insight into the way the field changes with depth 
into the cloud. The band D data most closely resemble the $850\,\mu$m data of 
\citet{kwon2018} that trace the coldest, densest material. Hence, we see that our 
data fill the gap in understanding at intermediate depths between the lowest 
density material and the highest.

We find a wide spread of polarization degree values (between 
$0\%$ and $\approx 15\%$), with median values for bands C and D at $7.5\%$ and $5.0\%$,
respectively. There is a clear tendency for lower polarization values to be
concentrated near the densest portions of the \roa/ core, with  median 
polarization values dropping to $5.1\%$ and $1.7\%$ for bands C and D, respectively, when
considering only points within $30\arcsec$ of the column density peak. 
This trend in polarization degree will be further explored in Section \ref{s:discussion}. 

\subsection{Column density and dust temperature maps}
\label{s:compdata}

The H$_2$ column density ($N$) and temperature ($T$) maps used in this work
(Figure \ref{f:ntmaps}) were derived from {\it Herschel} $70$, $100$ and 
$160\,\mu$m PACS data
\footnote{The list of PACS observing identification labels (OBSIDs) for these observations are the following: 1342205093, 1342205094, 1342227148, and 1342227149.}
\citep[Photodetector Array Camera and Spectrometer,][]{2010pacs} obtained from the {\it Herschel} Science Archive\footnote{\url{http://archives.esac.esa.int/hsa/whsa/}}. The $70$ and $100\,\mu$m  
maps were Gaussian-convolved to the same angular resolution of the $160\,\mu$m data ($11.4\arcsec$ FWHM)
and re-gridded to allow a pixel-by-pixel match. A modified thermal Spectral Energy Distribution (SED) fit was applied to each pixel using a fixed dust opacity
spectral index of $\beta = 1.62$ 
\citep{2014planckxi}. 

From each fit we obtain the dust temperature $T$ and the optical depth $\tau_{\lambda_o}$ at a reference wavelength chosen to be 
$\lambda_{o} = 250\,\mu$m. Following \citet{1983hildebrand}, $\tau_{\lambda_o}$ was then 
converted to H$_2$ column density ($N$) using the relation 
$\tau_{250} = \kappa_{250}\,\mu\,m_{\mathrm{H}}\,N$, where 
$\kappa_{250} = 0.1\,$cm$^2$g$^{-1}$ is the dust emissivity cross section per unit mass at $250\,\mu$m, $\mu = 2.8$ is the mean molecular weight, and $m_{\mathrm{H}}$
is the mass of the hydrogen atom.
The inclusion in our fits of more {\it Herschel} data at longer wavelengths (e.g., $250$, $350$, and $500\,\mu$m) did not cause a significant change in the values of
$T$ and $N$ obtained. Therefore, to preserve angular resolution in the final $N$ and $T$ maps, the results presented here employ only $70$, $100$ and 
$160\,\mu$m data.
It is important to point out that though we have fitted a single modified black-body to each pixel, 
in reality the total dust emission for any given pixel is comprised of contributions from several 
components along the line-of-sight (LOS) at different local temperatures. This notion will be 
further explored in 
Sections \ref{s:qualpspec} and \ref{s:modsimplemodel}. 
In order to avoid confusion between the local temperature and the temperature that is obtained by fitting the emission from a particular LOS, we will use different terms to refer to each of these quantities.  The latter quantity will be referred to as the ``LOS temperature'' or simply ``temperature", while the former quantity will be called the ``local temperature".


\subsection{Polarization spectra}
\label{s:pspecobs}

   \begin{figure*}[!t]
   \centering
   \includegraphics[width=0.49\textwidth]{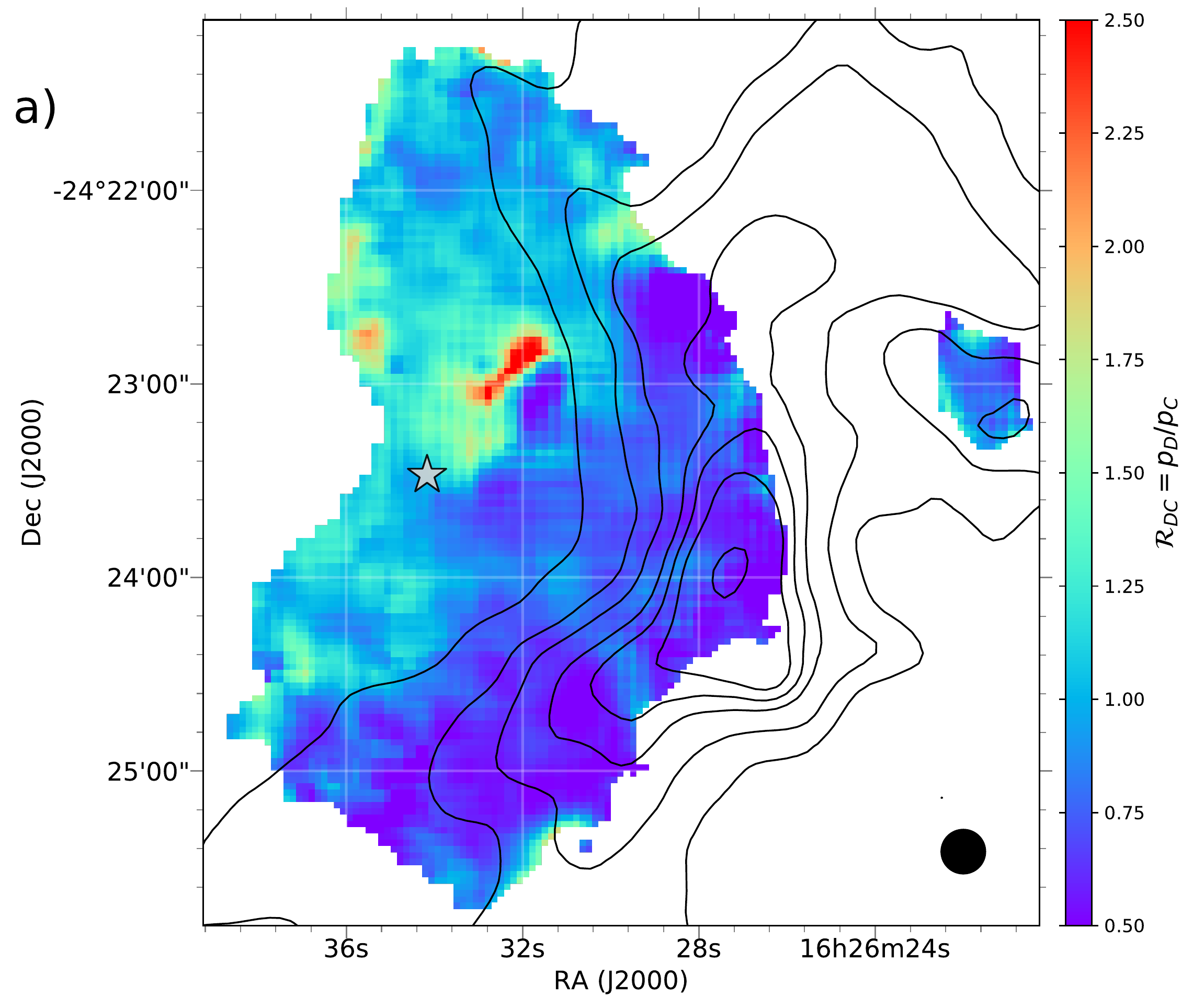} 
   \includegraphics[width=0.49\textwidth]{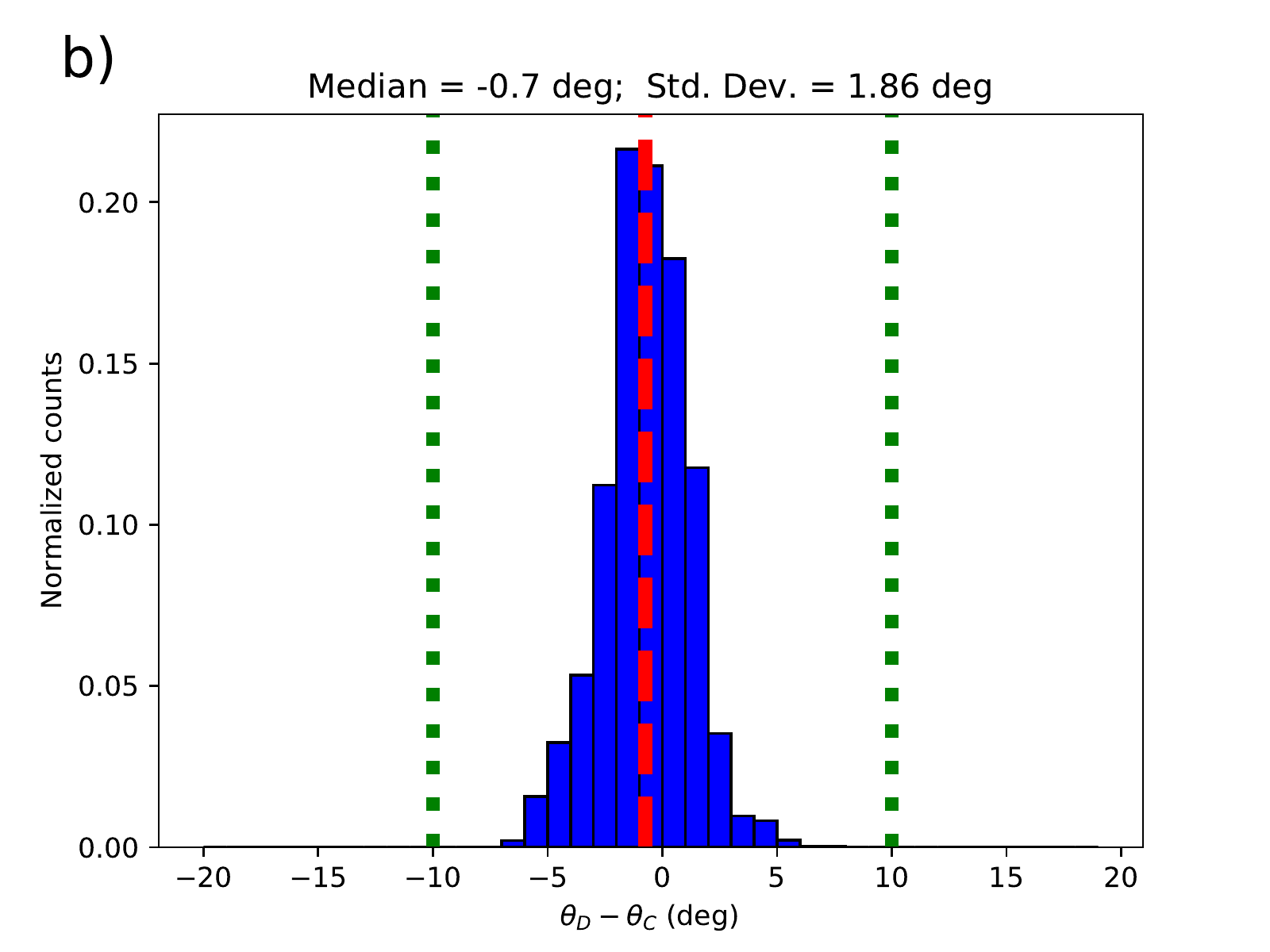}\\
   \includegraphics[width=0.49\textwidth]{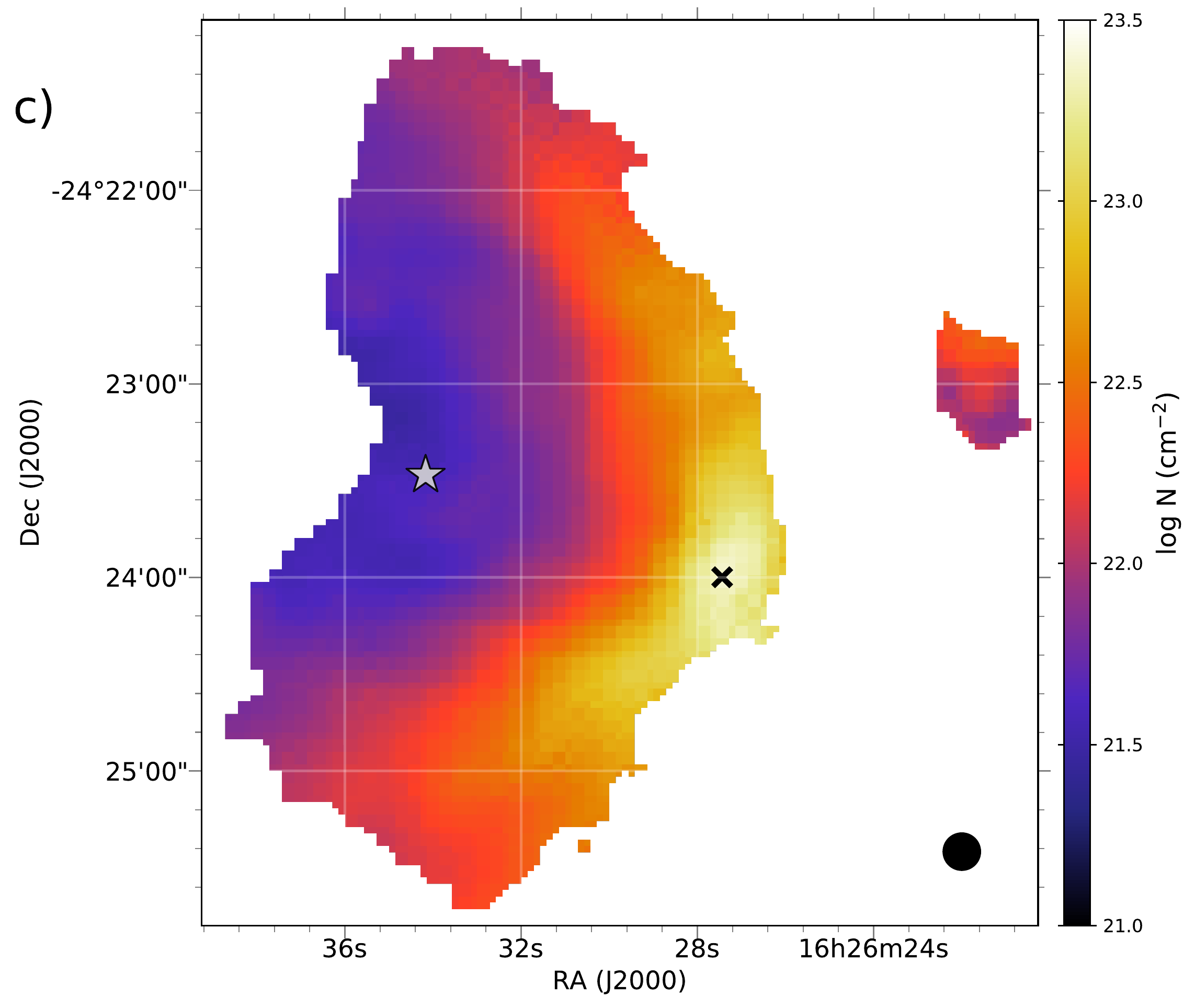}
   \includegraphics[width=0.49\textwidth]{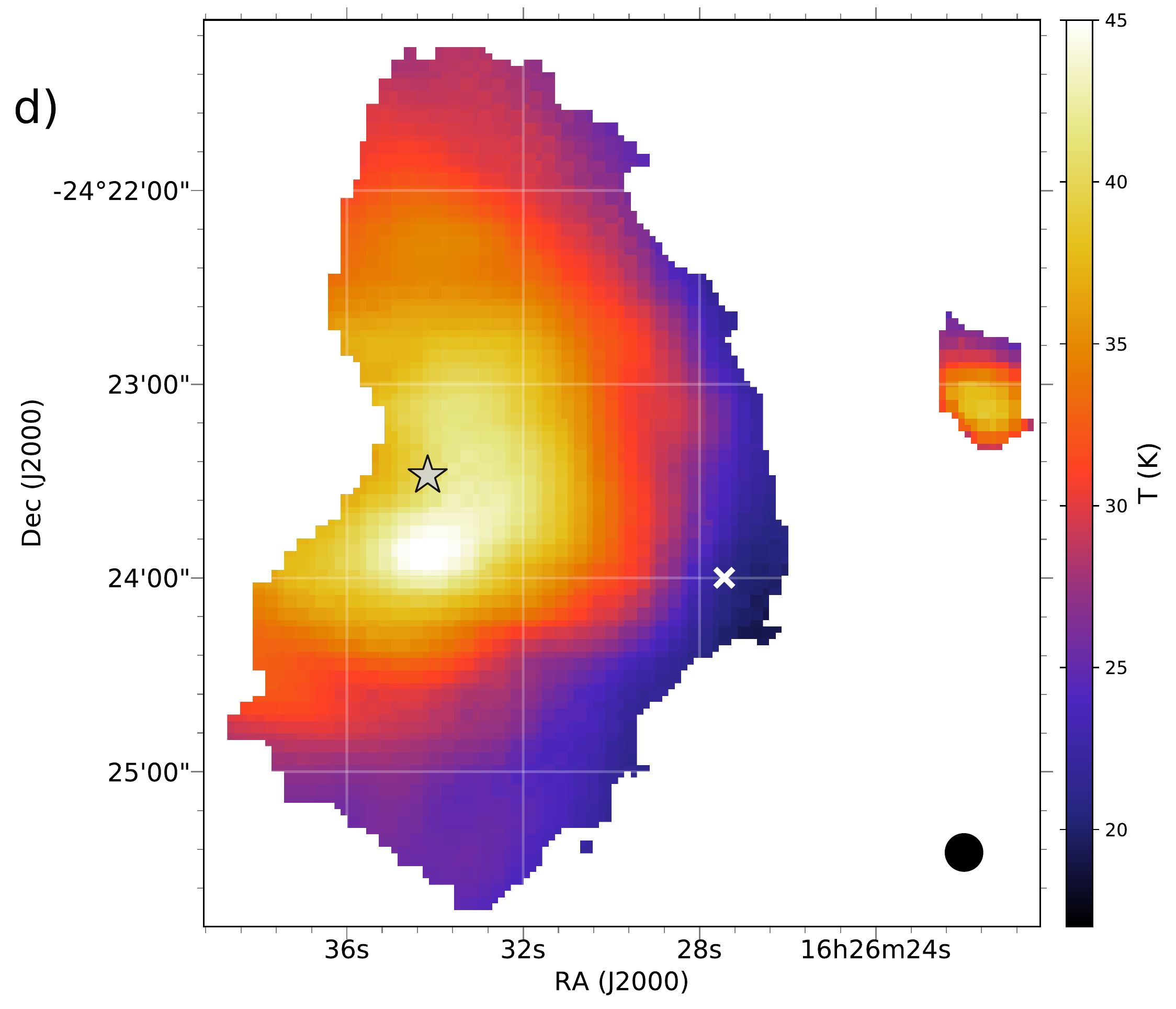}
   \caption{Polarization spectrum analysis of \roa/. (a) Map of 
   \rdc/$=p_{D}/p_{C}$, with $\mathrm{H}_2$ column density ($N$) contour levels distributed between  
   $1\times 10^{22}$ cm$^{-2}$ (lower level) and $2 \times 10^{23}$ cm$^{-2}$ (upper level). (b) Histogram of the 
   polarization angle difference between bands D and C ($\theta_{D} - \theta_{C}$); the red dashed
   vertical line indicates the median of the distribution 
   ($-0.7^\circ$), and the green dotted lines show the $10^\circ$ limit as a reference. Panels (c) and (d) show the same $N$ and $T$ maps as in Figure \ref{f:ntmaps}, but selecting only the map areas where \rdc/ data are available (compare with panel (a)) -- this is the ``polarization spectrum map area'', 
and it defines the input data used for the cloud core modeling in Section \ref{s:modsimplemodel}. The $\times$ symbol indicates the $N$ peak and the gray-colored star symbol indicates the position
   of Oph S1.
   The circle at the bottom right 
   of panels (a), (c), and (d) represents the corresponding beam size (FWHM). 
   }
         \label{f:pspec}
   \end{figure*}

To probe the slope of the far-infrared polarization spectrum across \roa/,
we define the ``polarization ratio'' as $\mathcal{R}_{DC} = p_{D}/p_{C}$. 
With this definition, positive and 
negative spectrum slopes are indicated by $\mathcal{R}_{DC} > 1$ and $\mathcal{R}_{DC} < 1$, respectively.
The first step in calculating $\mathcal{R}_{DC}$ is to re-generate the band C polarization maps 
using a merging Gaussian kernel
to match the band D beam size ($13.6\arcsec$ FWHM). Next, 
we calculate $\mathcal{R}_{DC}$ for each sky position. The resulting polarization
ratio map is shown in Figure \ref{f:pspec}a. Typically, investigations of polarization spectra of molecular cloud dust emission employ data 
masking to reject sky positions having
large differences in polarization angles across different wave-bands.  A typical threshold would be 
ten degrees (e.g., for our
case, points having
$|\theta_{D} - \theta_{C}| > 10^\circ$ would fail this cut).
As can be seen from the histogram in Figure \ref{f:pspec}b, 
we see a correlation in polarization angles between bands D and C that is very tight, 
with no corresponding detections in bands C and D having polarization angle 
differences larger than $10^\circ$.
Note that 
although the bands C and D polarization maps independently cover sky regions well beyond the area shown in 
Figure \ref{f:pspec}a, we select for analysis only those sky positions for which polarization detections are available in both bands.  
We will refer to this ensemble of sky positions as the ``polarization spectrum map area.'' 
This same criterion is applied to the column density and 
temperature maps (see Figures \ref{f:pspec}c and \ref{f:pspec}d).  
Thus, the analysis of column density and temperature presented below will
only make use of $N$ and $T$ data located within 
the polarization spectrum map area. 

A clear spatial trend is seen for the polarization ratio across \roa/: North-Eastern 
regions (including Oph S1) typically show \rdc/ $> 1$, while South-Western regions 
in general show \rdc/ $< 1$. Interestingly, these trends observed in \rdc/ can be correlated with trends observed in the column density and
temperature maps, which are shown in Figures \ref{f:pspec}c and \ref{f:pspec}d,
respectively. 
It appears that lower column density (and warmer) areas are typically associated with a positive polarization 
spectrum slope (\rdc/ $> 1$). Similarly, as one goes deeper into the cloud
(higher column density and colder regions), a negative polarization spectrum slope (\rdc/ $< 1$) 
becomes predominant.
For clarity, column density contours are also included in Figure \ref{f:pspec}a.

\subsection{Qualitative interpretation of the polarization ratio}
\label{s:qualpspec}

Let us picture the core 
of \roa/ as a dense structure embedded within a more diffuse molecular cloud, with the latter corresponding
to the extended molecular complex generally known as the 
\ro/ cloud. We reserve the term ``core" (or ``cloud core") to refer specifically 
to the denser structure, while the term ``ambient" medium
will refer to all the material along the LOS immediately outside the core, 
i.e., the ambient ISM corresponds to the more diffuse molecular cloud material.  
Finally, the term ``cloud" will refer to the combination of the core plus the ambient medium. 

As an initial approach to explain the negative (positive) correlation of \rdc/ with column density $N$ (temperature $T$), 
we will propose a qualitative picture based on a fall-off in
grain alignment efficiency as one goes from the core's outer edge to its inner higher density regions. 
As discussed in Section \ref{s:introduction}, such a fall-off is a prediction of RATs theory as well as being a 
leading candidate for explaining the widely observed anti-correlation between polarization fraction 
of emitted radiation and column density.  For purposes of the discussion in this Section, we will 
ignore other explanations for this anti-correlation (but we reconsider this in Section \ref{s:disc2}).  It is clear that grains in the core's more diffuse outer layers are more exposed to UV/optical radiation and 
therefore are naturally expected to be warmer when compared to those in the shielded high-density core's 
interior. In the context of RATs theory, the radiation not only heats the grains, but also increases
grains' efficiency for becoming aligned with respect to the magnetic field. In this case, grains 
located in the warm outer layers will be well aligned, with the alignment efficiency
gradually decreasing towards the core center.
From the observational perspective, there is strong evidence from numerous
studies that indeed the alignment efficiency seems to decrease at molecular cloud's 
interiors (see Section \ref{s:introduction}).
In their most general form, these trends in temperature and grain alignment
efficiency have been referred to as the extinction-temperature-alignment 
correlation \citep[ETAC,][]{2018ashton}.  We will use this term here.

We now argue that the ETAC provides a qualitative explanation for the anti-correlation between 
\rdc/ and $N$ that we have observed.  First, we introduce the term ``core limb'' to refer to the sky projection of the core's outer layers, analogously to the Sun's limb.
Next consider, from the observer's point of view, how the radiation observed at the core limb LOS differs from that observed for the LOS passing through the core's center.
The core limb LOS can be approximated as a single diffuse and warm component (neglecting for the moment the ambient medium component).
On the other hand, the core center LOS includes multiple components (outermost as well as inner core regions) having very different local temperatures and grain alignment efficiencies. 
For the core center LOS, we have a combination of: 
(1) well-aligned and warm dust particles in the core's outer layers, which favors an enhancement of the 
shorter wavelength polarization, $p_C$; with (2) poorly-aligned cold grains 
in the dense interior, which suppresses the longer wavelength polarized emission, $p_D$.
This combination should lead to a smaller \rdc/ value toward the center LOS in comparison with the core limb where we have a single diffuse component 
with well-aligned warm grains (no ETAC effect).  This is precisely the trend we observe in Figure \ref{f:pspec}a.  
Note that 
\citet{1999hildebrand} were the first to propose this same basic qualitative picture.  In that case,
the picture served to address the tension between the predicted flat spectra and the observed falling
spectra (see discussion in Section \ref{s:introduction}).

The qualitative scenario described above requires quantitative tests. To this end, as a sanity check on whether the ETAC can be considered a plausible 
explanation for why \rdc/ decreases as column density increases, we will now
compare our observations with a very crude, very simple toy 
model in which we assume that 
the ETAC is the only effect causing changes in grain alignment efficiency.
Our model is described in Section \ref{s:modsimplemodel}.
Note that the very simple model presented here is not intended to correspond to the 
core of \roa/ as a whole. Rather, because it uses only the $N$ and $T$ data
located within the polarization spectrum map area  (Figures \ref{f:pspec}c and \ref{f:pspec}d, respectively), it effectively represents only the Eastern side of
the core (which faces Oph S1).  This is all we can model, as this is the region for which we obtained corresponding polarization detections for both
bands C and D.

\subsection{Simple cloud model}
\label{s:modsimplemodel}

\subsubsection{Overview of the simple cloud model}
\label{s:modoverview}

\begin{figure}[!t]
   \centering
   \includegraphics[width=0.49\textwidth]{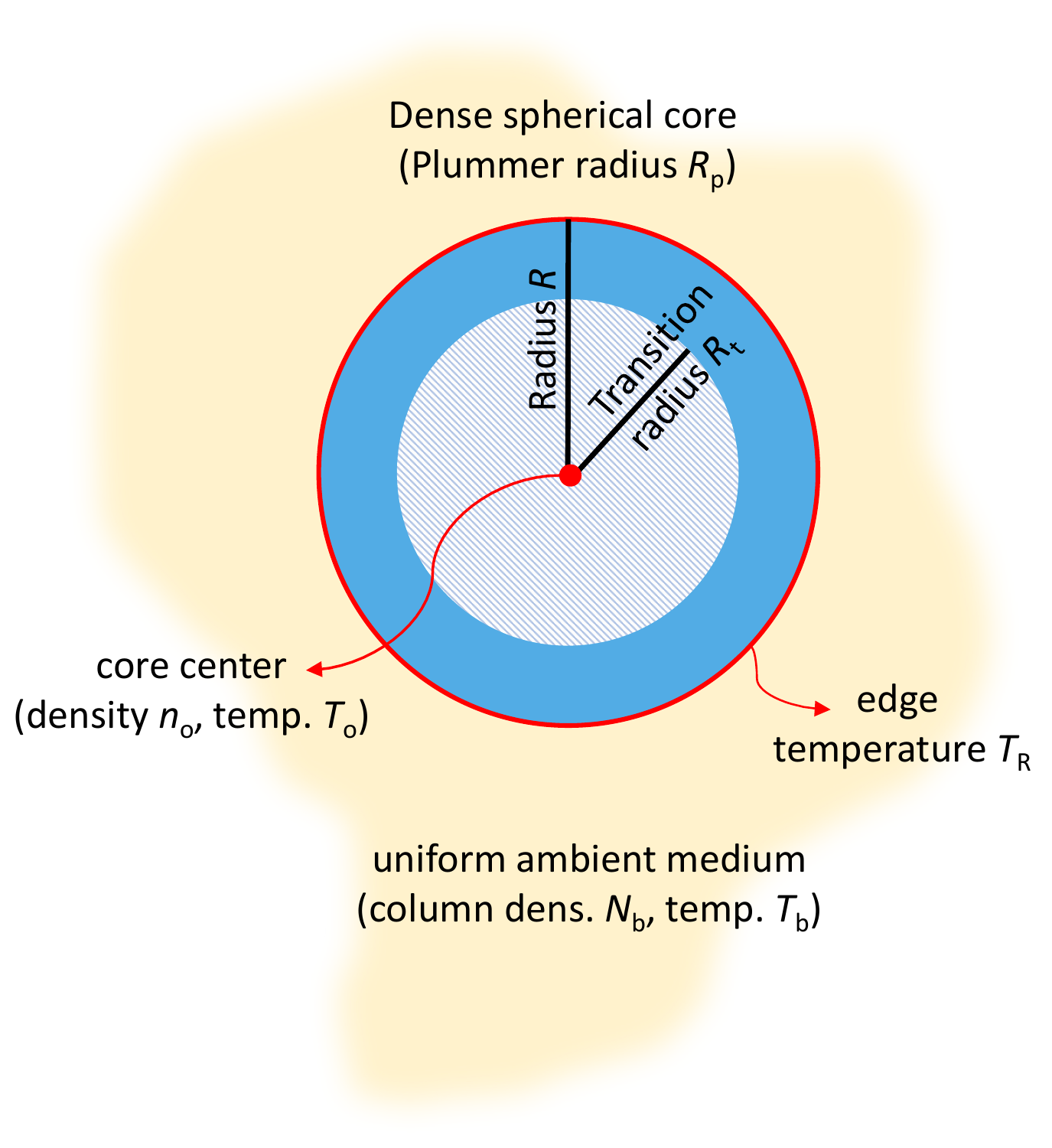} 
   \caption{Schematic representation of the simple cloud model described in
   Section \ref{s:modsimplemodel}. The seven parameters
   $R$, $N_{\mathrm{b}}$, $T_{\mathrm{b}}$, $n_{o}$, $R_{\mathrm{p}}$, $T_{o}$, and $T_{R}$
   are found by using the column density and temperature maps derived 
   from {\it Herschel} (Section \ref{s:compdata}), as described in 
   Sections \ref{s:modoverview} and \ref{s:modcloud}. The transition radius 
   $R_{t}$, described in Section \ref{s:modpol}, is introduced to test the hypothesis
   that grain alignment efficiency decreases toward the core center (i.e., the ETAC hypothesis):
   for $r < R_{t}$, the model assumes that grain particles are completely unaligned.
                 }
         \label{f:modcartoon}
   \end{figure}

The simple cloud model consists of a spherically symmetric 
core of radius $R$ embedded in a uniform ambient medium (or ``background") with H$_{2}$ column density $N_{\mathrm{b}}$ and temperature $T_{\mathrm{b}}$. 
We set the center of the spherical core to be located at the peak of the 
column density map (as indicated by the $\times$ symbol in 
Figures \ref{f:pspec}c and \ref{f:pspec}d). The core's H$_2$ number density profile 
is described by a Plummer sphere: 
$n(r) = n_{o}/(1 + r^{2}/R_{\mathrm{p}}^{2})^{5/2}$, where 
$r$ is the distance from the center, $n_{o}$ is the H$_2$ number density at the center,
and $R_{\mathrm{p}}$ is the Plummer radius. 
Similar density profiles (e.g., Plummer and ``Plummer-like") 
are frequently used
to describe molecular filaments \citep[e.g.,][]{2008nutter,arzou2011} and 
molecular cores \citep[e.g.,][]{2001whitworth,2002whitworth}.
For simplicity, the local temperature profile $T(r)$ is assumed to vary linearly between the 
core's center (where $T(0) = T_{o}$) and its edge ($T(R) = T_{R})$.
A schematic representation of the simple model, including the parameters
described above, is shown in Figure \ref{f:modcartoon}.
In the following analysis, we define $x$ to be the projected distance from the core 
center in the plane of the sky (assuming a cloud distance of $137\,$pc). 

The seven parameters $R$, $N_{\mathrm{b}}$, $T_{\mathrm{b}}$, $n_{o}$, $R_{\mathrm{p}}$, $T_{o}$, and $T_{R}$ completely
specify the wavelength-dependent total-intensity distributions in our model cloud.  Our strategy here 
is to first use the column density and temperature maps derived 
from {\it Herschel} (see Section \ref{s:compdata}) to determine these seven parameters before
proceeding to consider grain alignment, polarization, and the HAWC+ data.  Since the seven
parameters will depend directly on the $N$ and $T$ values derived from 
{\it Herschel}, it is important to note that these $N$ and $T$ values carry an intrinsic 
uncertainty related to the measurement errors and the inaccuracies in
the technique employed in Section \ref{s:compdata}.

Here is a step-by-step overview of our method:

\begin{itemize}
\item Step 1: From the {\it Herschel} $N$ and $T$ maps, we determine ``median NT curves'' 
that represent the input column 
density ($N$) and temperature ($T$) data as a function of $x$, the projected distance from the core center. 
Using these curves we determine 
the core radius
$R$ and the ambient medium parameters $N_{\mathrm{b}}$ and $T_{\mathrm{b}}$.

\item Step 2: We compute ambient-subtracted column density and temperature maps, 
allowing us to determine the local temperature at the core edge ($T_{R}$).

\item Step 3: We conduct a ``simulated observation'' of the simple model core + ambient medium, 
calculating ``output NT curves" that can be compared to the median NT curves. 
This procedure involves a flux integration along the LOS through the model cloud. 
The three remaining input parameters ($n_{o}$, $R_{\mathrm{p}}$, $T_{o}$) are determined
by minimizing the difference between the output NT curves and the median NT curves.

\item Step 4: We adopt a simple prescription for the dependence of grain alignment 
on core depth, and then integrate the polarized fluxes along the LOS. This allows us to 
estimate model curves of polarization degree (in bands C and D) and \rdc/ as functions
of column density, which can be directly compared with the
polarimetric observations.
\end{itemize}

Steps 1, 2, and 3 above are described in detail in Section \ref{s:modcloud}, while
Step 4 is elaborated in Section \ref{s:modpol}.

\subsubsection{Fixing the simple cloud model total intensity parameters}
\label{s:modcloud}

   \begin{figure*}[!t]
   \centering
   \includegraphics[width=0.49\textwidth]{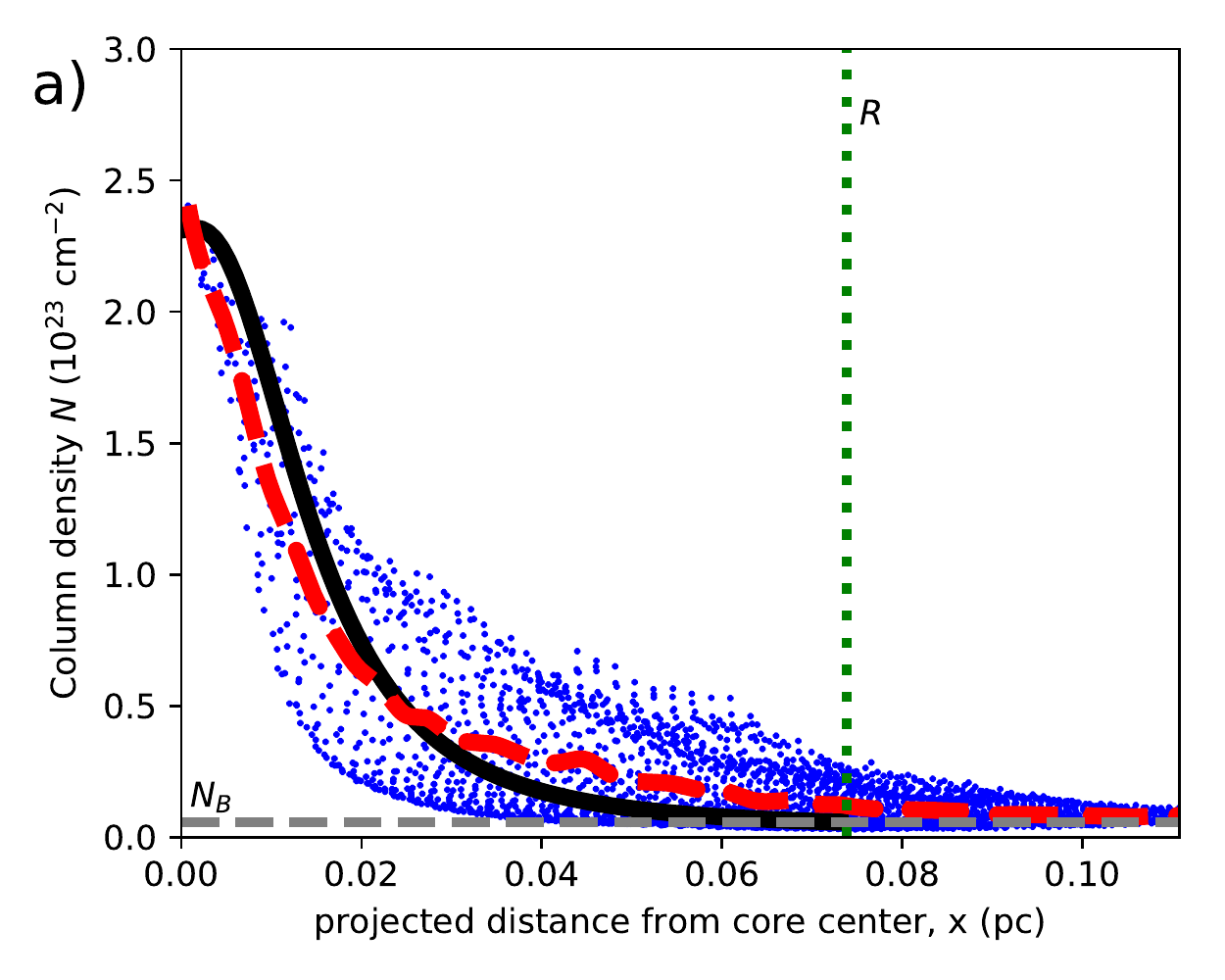} 
   \includegraphics[width=0.49\textwidth]{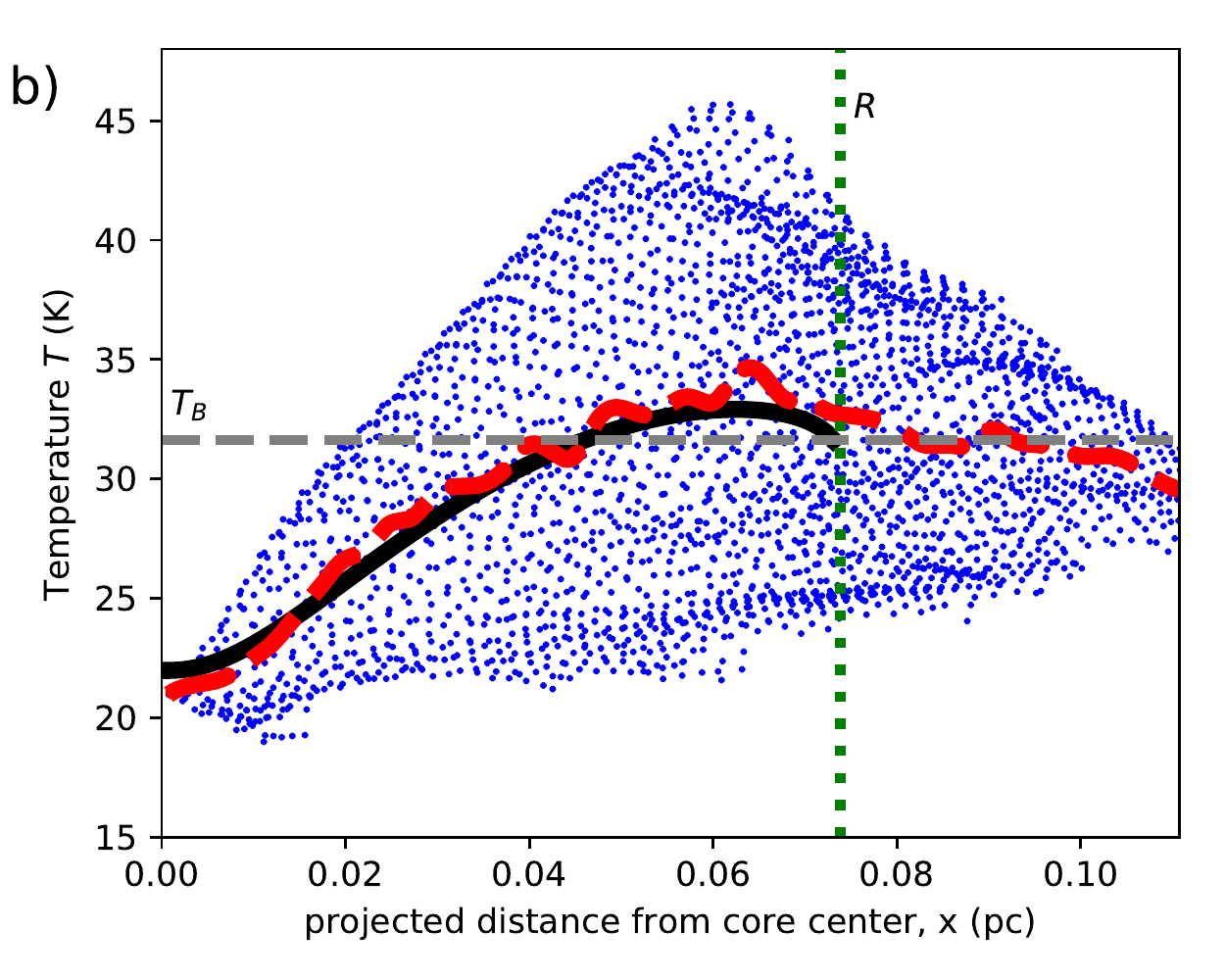} \\
   \includegraphics[width=0.49\textwidth]{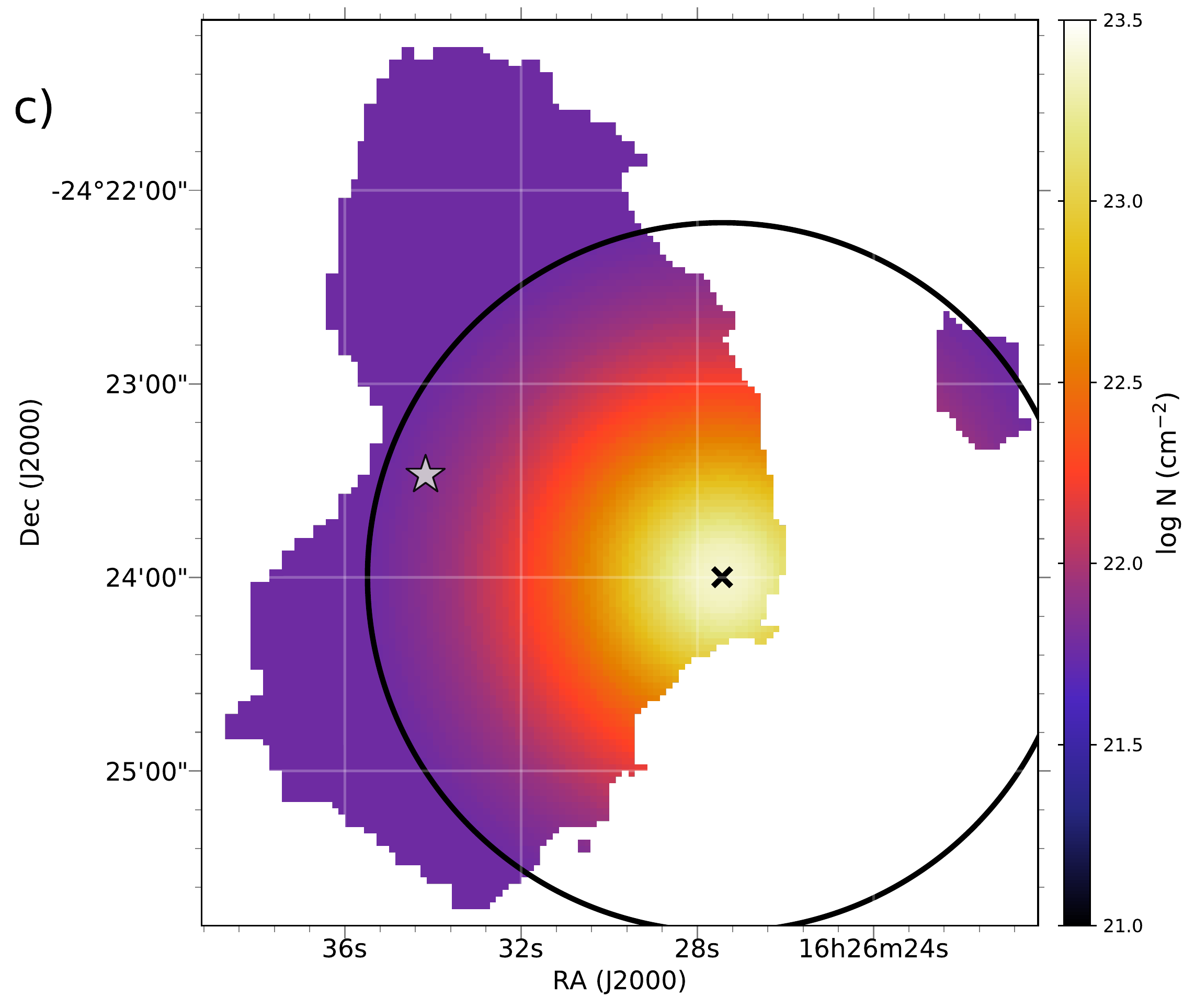}
   \includegraphics[width=0.49\textwidth]{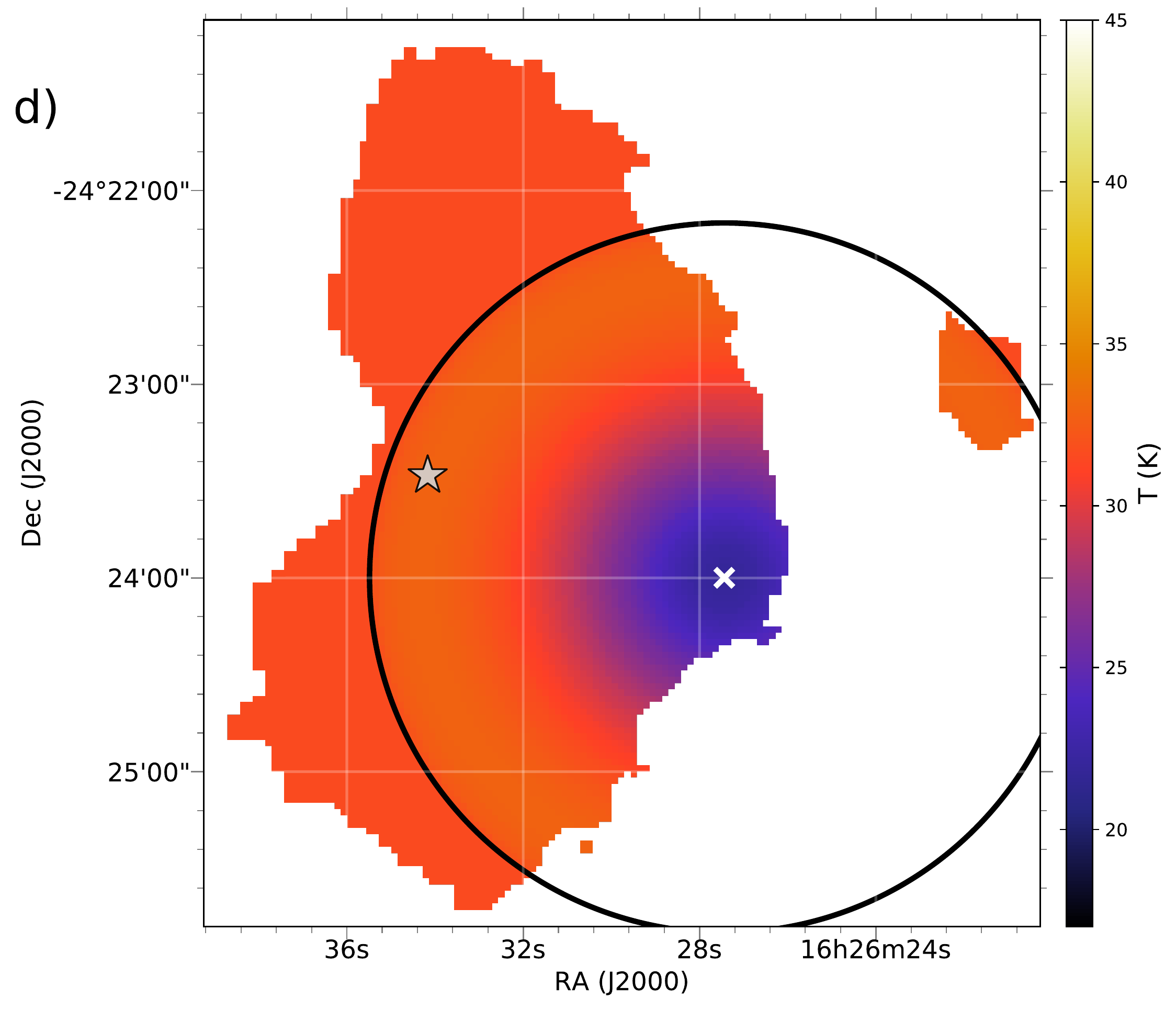}
   \caption{Spherical cloud core modeling of \roa/. Panels (a) and (b)
   show the $N$ and $T$ projected distance distributions (blue points),  
   using data shown in Figure \ref{f:pspec}c and d, respectively. 
   The red dashed lines are median NT curves using the 
   blue points as inputs. 
   The solid black lines are the model outputs for column density and 
   dust temperature after integrating the fluxes along the LOS and 
   simulating an observation of the spherical core (see Section 
   \ref{s:modcloud}). Gray horizontal dashed lines in panels (a)
   and (b) represent the column density ($N_{\mathrm{b}}$) and 
   temperature ($T_{\mathrm{b}}$) of the model 
   uniform ambient medium, respectively. Green vertical lines represent the core
   radius $R$. Mapping representations of the spherical core model (including the uniform ambient ISM)
   are shown in panels (c) (column density) and (d) (temperature). The $\times$
   symbol and the solid circle represent the center of the core and the 
   core radius $R$, respectively. The star symbol in panels (c) and (d)
   shows the corresponding position of Oph S1, and is given here just for reference 
   to compare with Figure \ref{f:pspec}.
                 }
         \label{f:modinputs}
   \end{figure*}

The determination of the seven total-intensity 
parameters for the simple model, following the steps listed in Section \ref{s:modoverview}, 
is carried out as follows:
First (Step 1 of Section \ref{s:modoverview}), we use the column density and temperature maps to build projected distance 
distributions of
$N$ and $T$, which are shown as blue points in 
Figures \ref{f:modinputs}a and b, respectively. 
As mentioned above (Section \ref{s:pspecobs}), in these projected distance distributions 
we include only data located within the polarization spectrum map area (these data are
shown in Figures \ref{f:pspec}c and \ref{f:pspec}d, respectively).
For both the $N$ and $T$ distributions as a function of $x$, median-binned curves 
are computed.  These are shown using red-dashed lines in \ref{f:modinputs}a and b. These curves are 
referred to as the ``median NT curves" and are denoted as $\overline{N}(x)$ and $\overline{T}(x)$. 
The goal is to find values for the seven total intensity parameters that reproduce these median NT curves 
as closely as possible.

We start by setting the limb of the model spherical core to be located at the
point where $\overline{N}(x)$ falls to 5\% of its peak
value.  This gives $R = 0.074\,$pc. 
To find $N_{\mathrm{b}}$ and $T_{\mathrm{b}}$, respectively, we take the median 
of all $N$ and $T$ values at $x > R$, restricting to the
polarization spectrum map area. The maximum $x$ value in this area is
$1.72R$, so $N_{\mathrm{b}}$ and $T_{\mathrm{b}}$ provide suitable 
estimates of the ambient column density and temperature immediately outside the 
cloud core: 
$N_{\mathrm{b}} = 5.7 \times 10^{21}\,$cm$^{-2}$ 
and 
$T_{\mathrm{b}} = 30.7\,$K. The 5\% choice described above for setting $R$ corresponds to
$\overline{N}(R)$ just 1$\sigma$ larger than $N_{\mathrm{b}}$,
where $\sigma$ is the standard deviation of the distribution of column density values for
$x > R$, which is a measure of the variation of the background column density that
has a mean value of $N_{\mathrm{b}}$.  (The numerical value of $\sigma$ is $5.0 \times 10^{21}\,$cm$^{-2}$.)
Thus, this 5\% choice appropriately limits the model core to a region where core emission
can be reasonably distinguished from background emission.

Keeping in mind that we now have four remaining total intensity parameters, 
$n_{o}$, $R_{\mathrm{p}}$, $T_{o}$, and $T_{R}$, we move to Step 2 of 
Section \ref{s:modoverview}.  We find that one of these parameters, $T_{R}$, can be 
directly inferred from the {\it Herschel}-derived temperature map in a
straightforward way.  This is because, from the observer's perspective, the core limb 
LOS involves the superposition of only two cloud components: the uniform ambient medium
and the layer of the core for which $r = R$.
Accordingly, from $N_{\mathrm{b}}$ and $T_{\mathrm{b}}$ we compute the ambient ISM  
flux values $I_{\lambda,b}$ for each {\it Herschel} wavelength $\lambda$, using the same modified 
blackbody SED and scaling relations described in Section \ref{s:compdata}.
We then remove this ambient contribution by subtracting these wavelength-dependent flux values
from each {\it Herschel} map. 
Next, we compute ambient-subtracted column density and 
temperature maps using a procedure identical to the one described in Section \ref{s:compdata}. From the 
ambient-subtracted temperature map, we find $T_{R} = 38.9\,$K by taking the median 
temperature
value at $x = R$. For completeness, we note that the median ambient-subtracted column density 
at $x = R$ is $2.0 \times 10^{21}\,$cm$^{-2}$.

A key ingredient in our method for setting the values for the last three parameters (Step 3 of Section \ref{s:modoverview}) is 
our procedure for conducting a ``simulated {\it Herschel} observation" of the model cloud,
yielding column density and temperature profiles (``output NT curves") that can 
be compared to the
median NT curves ($\overline{N}(x)$ and $\overline{T}(x)$) of Figures \ref{f:modinputs}a and b. 
First, we choose 100 values of $x$ that are uniformly distributed between $x = 0$ and $x = R$, 
and then for each combination of one of these $x$ values and one choice of {\it Herschel}-band,
we integrate the {\it Herschel}-band emission at each core radius $r$ along the line-of-sight 
to find the flux $I_{\lambda}(x)$: 

\begin{equation}
\label{e:tflux}
I_{\lambda} (x) = \kappa_{250}\, \mu\, m_{\mathrm{H}}\,\left(\frac{\lambda_{o}}{\lambda}\right)^{\beta} \int n(r)\,B_{\lambda}(T(r))\,ds + I_{\lambda,b}
\end{equation}

\noindent In the expression above, $\lambda$ is the wavelength corresponding to the {\it Herschel}-band,
$s$ represents distance along the 
LOS, $B_{\lambda}$ is the {\it Planck} function,
and the parameters $\kappa_{250}$, $\mu$, $m_{\mathrm{H}}$, and $\beta$ were defined 
in Section \ref{s:compdata}. 
Notice that, in contrast with our earlier treatment that is valid for arbitrary optical 
depth (Section \ref{s:compdata}), Equation \ref{e:tflux} relies on the optically thin approximation.  
As discussed in Section \ref{s:disc3}, this introduces only a modest level of 
error in the final calculated polarization ratios.

Secondly, we use the simulated observed fluxes
$I_{\lambda}(x)$ to compute model column density $N_{m}(x)$ and 
temperature $T_{m}(x)$ values for each value of $x$, 
using the procedure described in Section \ref{s:compdata}
(i.e., fitting a modified black-body function to the fluxes). 
Obviously this entire ``simulated observation" procedure requires values for all
seven total intensity parameters, including the three that have until now
remained unconstrained ($n_{o}$, $R_{\mathrm{p}}$, and $T_{o}$). 
For a given set of parameters, we can compute 
curves of $N_{m}(x)$ and $T_{m}(x)$ (as exemplified in Figures \ref{f:modinputs}a and \ref{f:modinputs}b
by the thick black curves), and compare them to the curves of $\overline{N}(x)$ and 
$\overline{T}(x)$ (red-dashed lines in the same figures). 

To determine the best values for $n_{o}$, $R_{\mathrm{p}}$, and $T_{o}$, 
we ran the simulated model observation multiple times, varying these three parameters in each run, 
and searched for the set that minimizes the difference between the 
output NT curves ($N_{m}(x)$ and $T_{m}(x)$) and the median NT curves ($\overline{N}(x)$ and $\overline{T}(x)$).
For each run, we calculated the quantity 
$\Delta_{NT} = \sum_{x} (N_{m}(x) - \overline{N}(x))^2 \times \sum_{x} (T_{m}(x) - \overline{T}(x))^2$, 
which can be understood as the combined summed square difference between the column density and 
temperature curves. The quantity $\Delta_{NT}$ was minimized for the following best-fit values:
$n_{o} = 6.4 \times 10^{6}\,$cm$^{-3}$, 
$R_{\mathrm{p}} = 0.244\,R$, and 
$T_{o} = 13.9\,\mathrm{K}$.  
The adopted values for all seven total intensity parameters are collected in Table \ref{t:tab}. 
The thick black curves in Figures \ref{f:modinputs}a and b show $N_{m}(x)$ and $T_{m}(x)$ 
as computed from these adopted parameters.


\begin{deluxetable*}{ccc}[t!]
\tablecaption{Parameters of the simple spherically symmetric cloud-core model \label{t:tab}}
\tablecolumns{3}
\tablewidth{0pt}
\tablehead{
\colhead{Parameter} & \colhead{Description} & \colhead{Determined value}
}
\startdata
$R$ & Core radius & $0.074\,$pc   \\
$N_{\mathrm{b}}$ & H$_2$ ambient column density & $5.7 \times 10^{21}\,$cm$^{-2}$ \\
$T_{\mathrm{b}}$ & Ambient dust temperature & $30.7\,$K   \\
$T_{R}$ & Local dust temperature at the core edge ($r = R$) & $38.9\,$K   \\
$n_{o}$ & H$_2$ number density at the core center ($r = 0$) & $6.4 \times 10^{6}\,$ cm$^{-3}$ \\
$R_{\mathrm{p}}$ & Core Plummer radius & $0.244\,R$   \\
$T_{o}$ & Local dust temperature at the core center ($r = 0$) & $13.9\,\mathrm{K}$  \\
$R_{t}$ & Polarization transition radius & $0.6\,R$ \\
\enddata
\end{deluxetable*}

Lastly, using these final $N_{m}(x)$ and $T_{m}(x)$ curves, Figures \ref{f:modinputs}c and d
show map representations of the column density and temperature profiles, respectively, 
for our simple model. Given the limitations of our spherical-core approximation
for \roa/, these maps provide something close to the best possible representation of the real cloud
based on the simple model and the inputs from the {\it Herschel} observations.
Comparing Figures \ref{f:pspec}c and d with Figures \ref{f:modinputs}c 
and d, we note that some of the very general features of the column density and 
temperature maps are well reproduced by the simple model, e.g., the low temperatures
and high column densities near the core center, as well as the lower column densities and 
warmer temperatures surrounding the core. However, there are many obvious differences between
the real and model maps. 
Our simple approach provides for an initial sanity check (as discussed in 
Section \ref{s:qualpspec} and also below), but in Section \ref{s:disc3} we discuss ways in which the 
model could be improved in future investigations.

\subsubsection{Polarization degree and polarization ratio for the simple cloud model}
\label{s:modpol}

In the paradigm described in Section \ref{s:qualpspec}, based on ETAC, dust grains have gradually 
decreasing alignment efficiency going from the diffuse outskirts of the core 
towards its dense interior. 
For the purpose of the very simple toy model developed in this work,
we choose the simplest grain alignment prescription that is consistent with ETAC:
we assume that at very high densities, interior to a certain cutoff radius,
dust grains are completely unaligned, so that the polarization detected 
toward the central parts of the dense core actually originates in the core's
more diffuse outer layers (i.e., there is no internal heating from embedded sources).
Accordingly, we define a cutoff ``transition radius" $R_{t}$ interior to which the alignment 
efficiency is set to zero. Given this assumption, the goal here 
is to calculate the expected values of \rdc/ from the model cloud, in order to compare 
with the observations (following Step 4 described in Section \ref{s:modoverview}). 
For each sky-projected distance $x$ from the 
core center, we use the definition of  polarization degree: 
$p_{\lambda}(x) = P_{\lambda}(x)/I_{\lambda}(x)$, where $P_{\lambda}(x)$ is the 
polarized flux (see below) and $I_{\lambda}(x)$ is the total flux
(e.g., as might be computed using Equation \ref{e:tflux}). To account for the 
effect of a cutoff radius for the polarization efficiency, 
we calculate the polarized flux according to the following prescription:

\begin{equation}
\begin{aligned}
P_{\lambda}(x) = p_{\lambda,o} I'_{\lambda}(x) \ , \\
\end{aligned}
\end{equation}

\noindent where $I'_{\lambda}(x)$ is calculated as in Equation \ref{e:tflux}, but excluding the
region between $s=-\sqrt{R_t^2-x^2}$ and $s=+\sqrt{R_t^2-x^2}$ from the integral. 
Parameter $p_{\lambda,o}$ represents the ambient 
polarization efficiency outside the volume defined by the transition radius. For simplicity, 
it is assumed to be spatially uniform. For each band, $p_{\lambda,o}$ is found by taking the median polarization degree 
for all polarization detections
within the polarization spectra map area having $x > R_{t}$.

Using this simple model, we calculate the values of polarization degree in bands D and C
as a function of $x$: $p_{D,m}(x)$ and $p_{C,m}(x)$, respectively.
In addition, the model polarization ratio \rdcm/$(x)= p_{D,m}(x)/p_{C,m}(x)$
is computed. 
By combining with the  $N_{m}(x)$ and $T_{m}(x)$ curves obtained in Section
\ref{s:modcloud}, we can find model-estimated curves of 
any of $p_{D,m}$, $p_{C,m}$ and \rdcm/ as a function of 
either column density $N_{m}$ or temperature $T_{m}$.

Note that this simple polarization model for the cloud has only one free parameter, 
the transition radius $R_{t}$.
For the purpose of this analysis we have chosen three values: 
$R_{t} = 0.3\,R$, $0.6\,R$, and $0.9\,R$
(these values were selected since they encapsulate the full 
parameter space of the observations, which we will describe 
in more detail in Section \ref{s:disc2}).
Comparisons between observed and model curves of polarization degree and polarization 
ratio are shown in Figure \ref{f:pspecgraphs}.
In each panel of Figure \ref{f:pspecgraphs}, 
the yellow-blue-red colored background represents 
2D histograms of the HAWC\plus/ observations: 
$p_{C}$ vs. $\log{N}$ in panel (a), 
$p_{D}$ vs. $\log{N}$ in panel (b), \rdc/ vs. $\log{N}$ in panel (c), 
and \rdc/ vs. $T$ in panel (d).
The colored curves represent the model $p_{C,m}$, $p_{D,m}$ and \rdcm/ as a function of $\log{N_{m}}$ 
(Figures \ref{f:pspecgraphs}a, \ref{f:pspecgraphs}b, and \ref{f:pspecgraphs}c, respectively) and 
\rdcm/ as a function of $T_{m}$ (Figure \ref{f:pspecgraphs}d)
using the three chosen values of $R_{t}$, as explained above.
These graphs are interpreted and discussed in Section \ref{s:discussion}.

\section{Discussion}
\label{s:discussion}

\subsection{Comparison between the observed far-IR polarization spectrum \\ 
and physical dust model predictions}
\label{s:disc1}

\begin{figure*}[!t]
   \centering
   \includegraphics[width=0.495\textwidth]{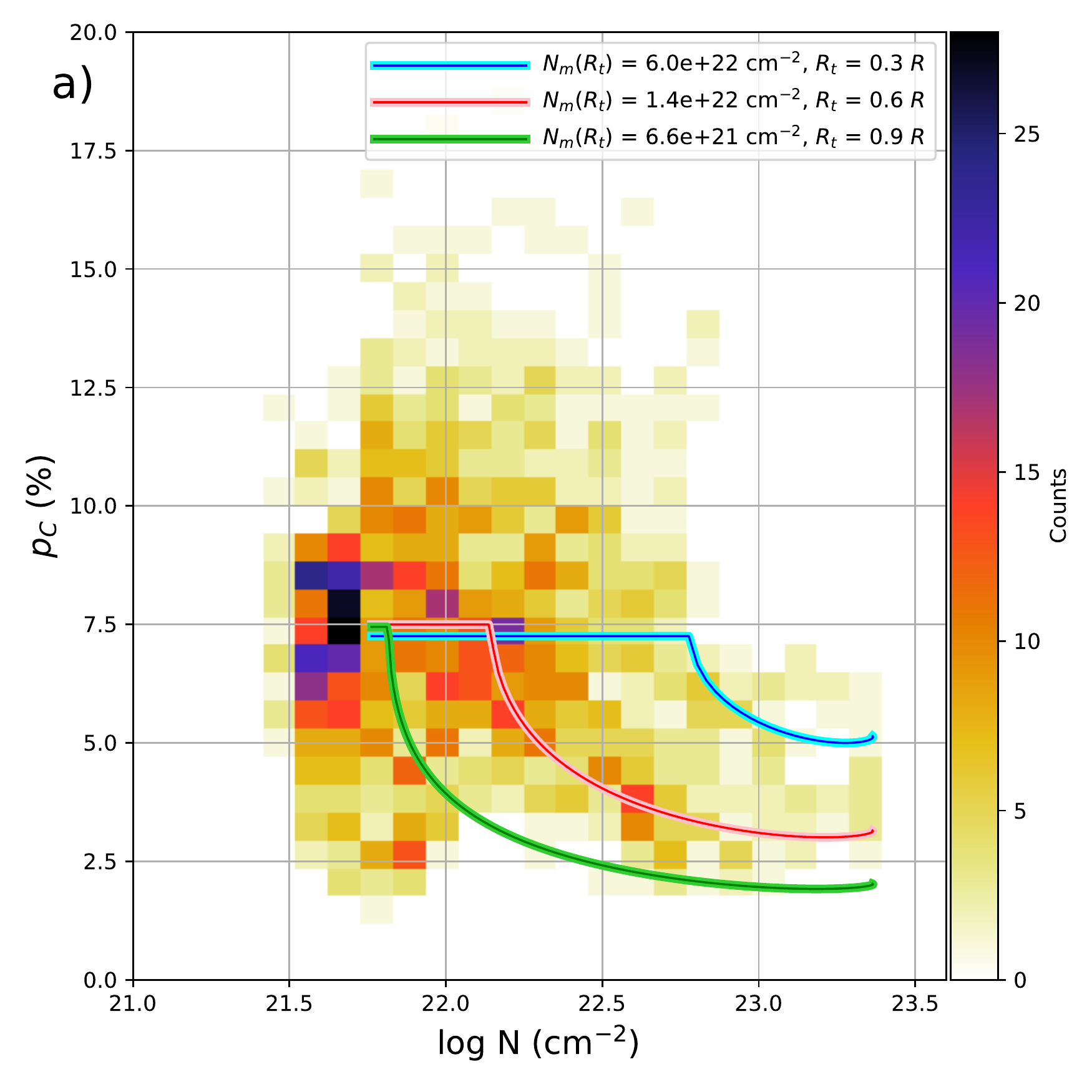} 
   \includegraphics[width=0.495\textwidth]{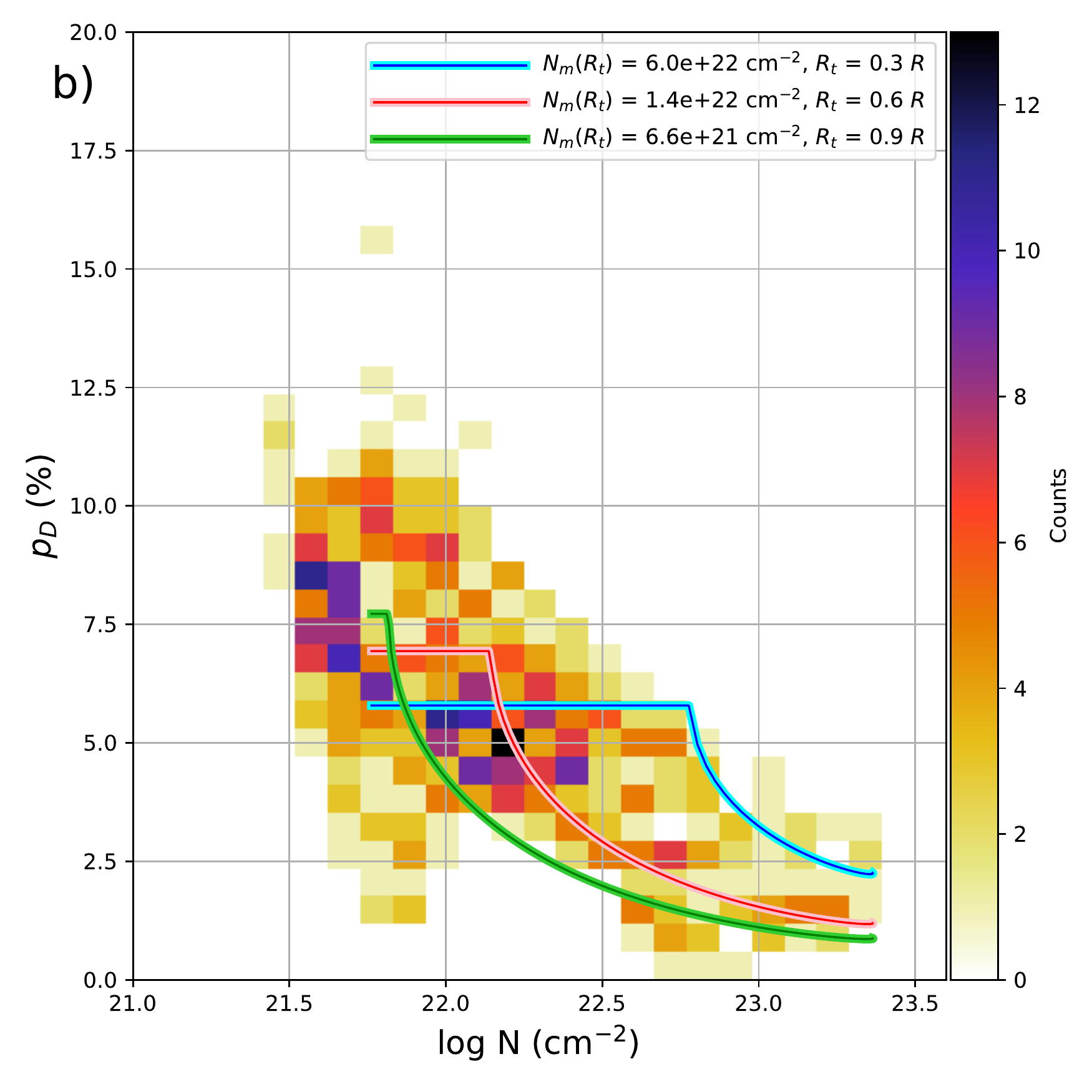} \\
   \includegraphics[width=0.495\textwidth]{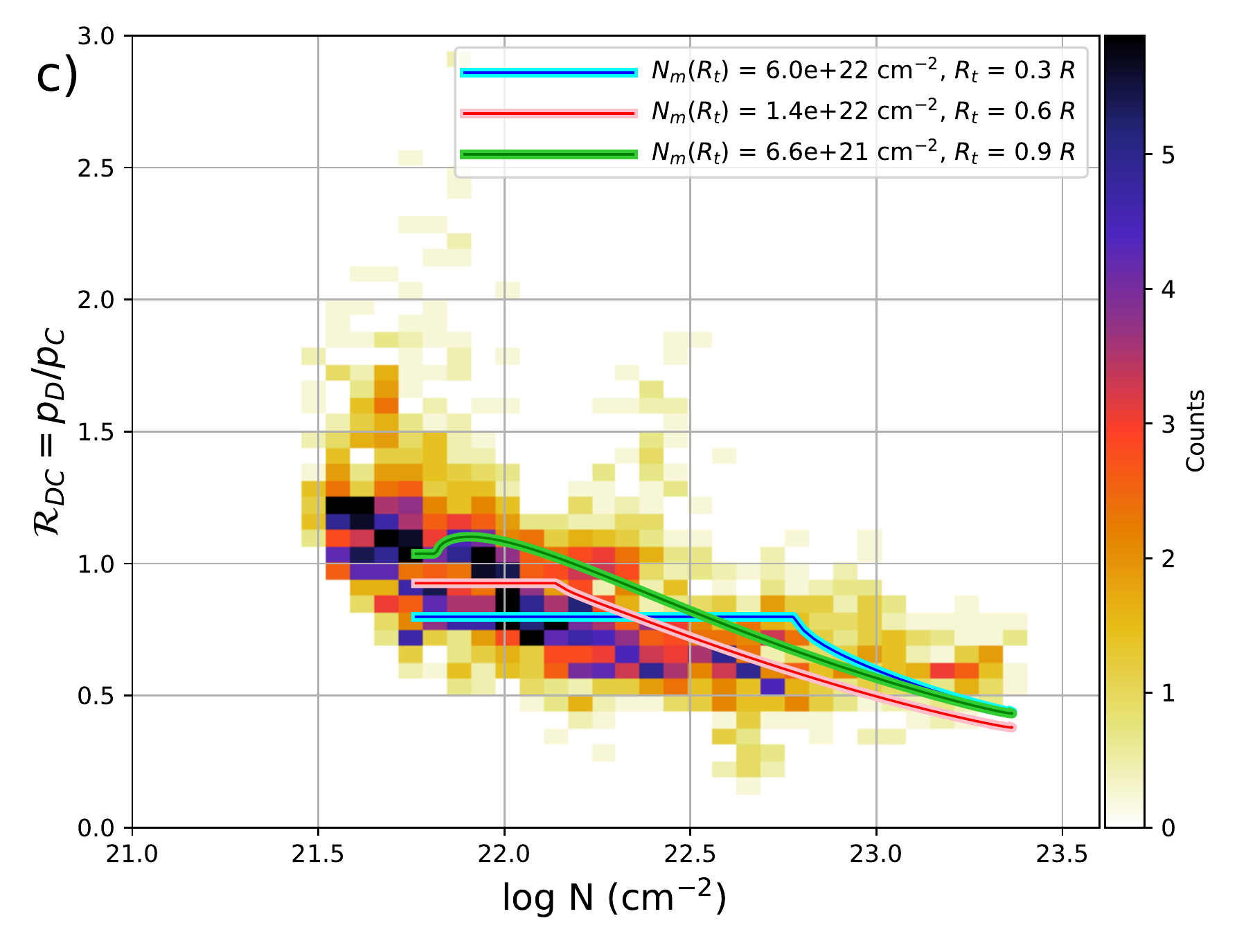}
   \includegraphics[width=0.495\textwidth]{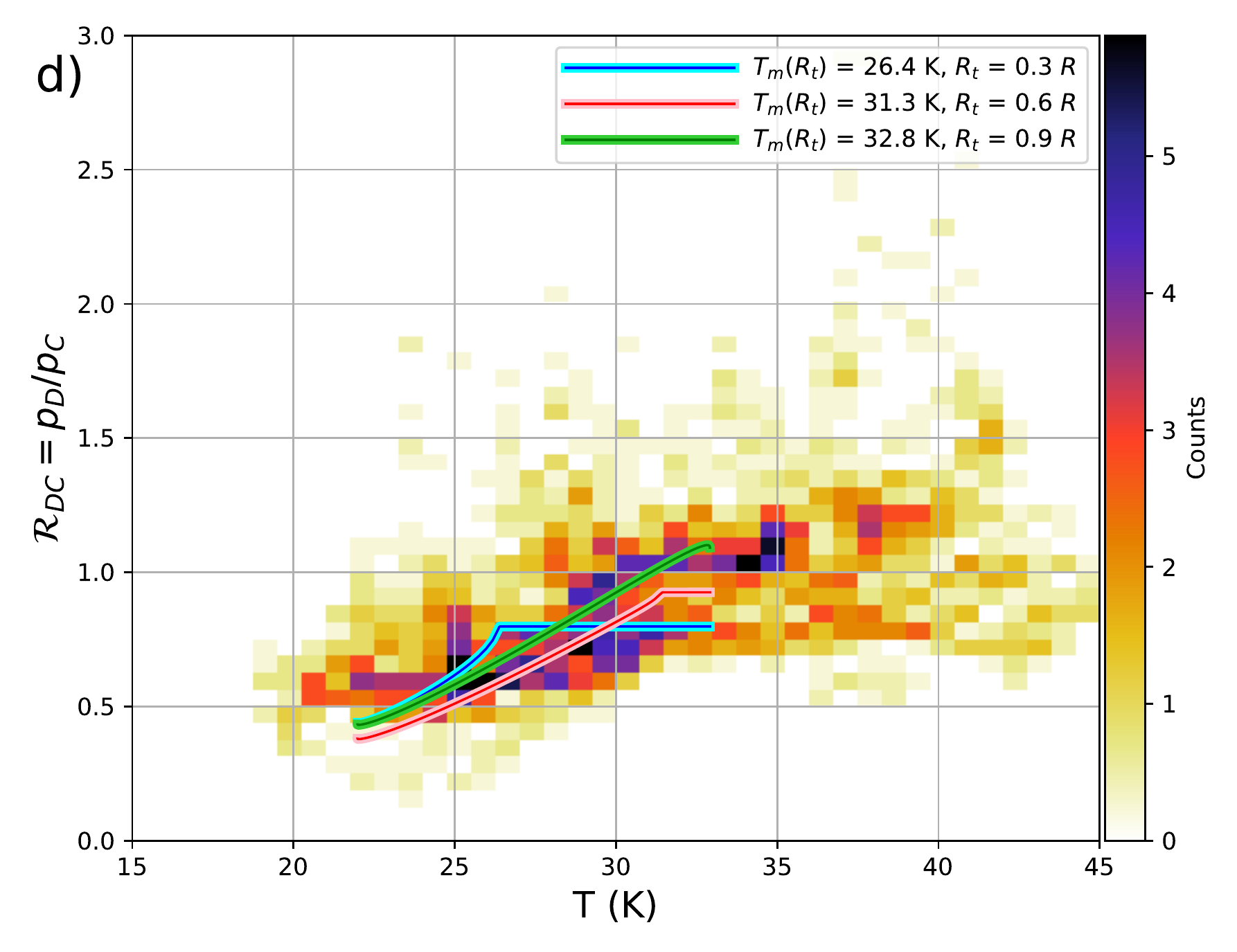}
   \caption{Comparison of the HAWC\plus/ polarization data (colored 2D 
   histograms), with the predictions of the simple cloud model 
   of \roa/ (colored curves): 
   (a) $p_{D}$ vs. $\log N$; 
   (b) $p_{C}$ vs. $\log N$; 
   (c) \rdc/ $=p_{D}/p_{C}$ vs. $\log N$; 
   (d) \rdc/ $=p_{D}/p_{C}$ vs. $T$.
   The blue, red and green solid lines represent three choices
   of the transition radius $R_t$: $0.3 R$, $0.6 R$, and $0.9 R$, 
   respectively. 
    The legends in each panel indicate the model column densities (panels a, b, and c) 
   and model LOS temperatures (panel d) associated with each of these $R_t$ values.
              }
         \label{f:pspecgraphs}
   \end{figure*}

The \rdc/ map presented in Section \ref{s:pspecobs} suggests correlations 
of the polarization spectrum slope with  column density $N$. 
In 
the 2D histogram of the observed \rdc/ as a function of column density 
(Figure \ref{f:pspecgraphs}c),
such a correlation is clearly seen: \rdc/ goes from $>1$ to $<1$ as 
$N$ increases. 
Given that $N$ and $T$ are highly anti-correlated within the polarization spectrum 
map area (compare Figures \ref{f:pspec}c and \ref{f:pspec}d) 
a correlation of \rdc/ with $T$ is also expected. In fact,
viewing \rdc/ as a function of temperature (Figure 
\ref{f:pspecgraphs}d), \rdc/ changes from $<1$ to $>1$ as 
$T$ increases.
Previous observations of the far-infrared polarization spectrum toward star-forming
clouds have detected negative polarization spectrum slopes 
\citep{2002vai,2008vai,2012vai,2013zeng},
but no clear systematic dependence on column density (or temperature) has been reported previously.
In particular, the positive polarization spectrum slopes we observe
toward the more diffuse areas surrounding the core provide a connection 
with the grain alignment models of \citet{bethell2007}, \citet{draine2009}, and \citet{2018guillet}, 
all of which predict positive slopes for this wavelength range.

An approximate quantitative comparison between the positive polarization 
spectrum slopes observed toward \roa/
and the predicted \rdc/ values from the models available in the literature can be made. 
For instance, although the modeled polarization spectra from \citet{bethell2007} 
is averaged over a wide range of column densities, their mean $N$ value lies 
between $\approx 10^{21.5}\,$cm$^{-2}$ and 
$\approx 10^{22.0}\,$cm$^{-2}$. This range is similar to the lower column density 
coverage of the HAWC\plus/ observations toward \roa/ (see Figure \ref{f:pspecgraphs}c).
For \roa/, we find a mean \rdc/ value of $1.1$ within this same range of column
densities. From the models by \citet{bethell2007}, we estimate predicted 
\rdc/ values between $1.75$ and $2.0$ based on their Figure 13. Notice, however, 
that the dust temperatures assumed by \citet{bethell2007} (in the range $\approx 5 - 17\,$K)
are significantly colder than the temperatures in the positive-slope polarization 
spectrum region of \roa/ (between $\approx30\,$K and $\approx45\,$K, near Oph S1 --
see Figure \ref{f:pspec}d).

\citet{2018guillet} also presented predicted polarization spectrum 
curves ($p$ vs. $\lambda$), although 
for a somewhat lower column density regime, between $\approx 10^{21.0}\,$cm$^{-2}$ and 
$\approx 10^{21.5}\,$cm$^{-2}$ (translucent clouds). They point out that there is 
a strong correlation between the dust temperature and the intensity of the interstellar
radiation field (ISRF). Moreover, as shown by their Figure 15, the polarization spectrum 
curves are significantly affected by the ISRF level. In the range of models presented 
by \citet{2018guillet} with different ISRF intensities, the ones with higher ISRF 
levels show \rdc/$\,\approx 1.1$, while models with lower ISRF intensities show 
\rdc/$\,\approx 20.0$. This suggests that the \rdc/ parameter within the lower column density 
regime is expected to be significantly affected by the level of exposure to 
radiation (and consequently, by the dust temperature).
Given the proximity of the lower-density positive-slope regions of \roa/ to
Oph S1, it is plausible to speculate that the somewhat lower \rdc/
values as compared to the models might be due to the strong exposure to 
radiation (and warmer dust temperatures) 
in this area. A more accurate comparison between models and observations of 
the far-IR polarization spectrum slope for lower density regimes requires an
accurate treatment of the ISRF, which is beyond the scope of this work.
For completeness, it is worth pointing out that although the \citet{draine2009} 
models probe a very different regime (the diffuse ISM), they 
predict \rdc/ values between $1.4$ and $2.6$, which are also slightly 
higher than the mean values found in \roa/ for the lower column density
areas.




\subsection{Analysis of the results from the simple spherical cloud model of \roa/}
\label{s:disc2}

The correlations between \rdc/ and the column density, 
found in the observational data, motivated the 
development of our simple cloud model with the goal of investigating whether ETAC can 
explain the correlations or not. Below we discuss the comparison between 
the observational data and the curves generated from this simple model (see 
Figure \ref{f:pspecgraphs}).
In Figures \ref{f:pspecgraphs}a and \ref{f:pspecgraphs}b, the 
observations exhibit a decrease of polarization degree as column density increases. 
This trend is commonly observed in molecular clouds 
\citep{goodman1995,gerakines1995,matthews2002,whittet2008,chapman2011,cashman2014,alves2014,jones2015,2016fissel}.
Similarly, the simple model curves shown in blue, red, and green also exhibit 
a decreasing trend in the $p$ vs. $N$ graphs (Figures 
\ref{f:pspecgraphs}a and \ref{f:pspecgraphs}b). 
As noted above, we 
have chosen a set of three values for the free parameter $R_{t}$: 
$0.3\,R$, $0.6\,R$, and $0.9\,R$. The reason for these particular choices is that
the extreme values ($0.3\,R$ and $0.9\,R$, green and blue curves, respectively) 
represent an approximate ``boundary'' for the spread in the observational data points of 
Figure \ref{f:pspecgraphs} panels (a) and (b). Therefore, choices outside 
this range are probably poor fits to the data. In this sense the midpoint
choice ($R_{t} = 0.6\,R$, red line, and also given in Table \ref{t:tab}) is a reasonable best fit to the observed dependence of $p_C$ and $p_D$ on $N$.
Note that we can directly associate each value of the transition radius with 
a corresponding column density $N_{m}(R_t)$ 
along the LOS.
For $R_{t} = 0.6\,R$ we find  $N_{m}(R_t) = 1.4 \times 10^{22}\,$cm$^{-2}$. For column 
densities larger than this value, grains are no longer aligned.

Within the range of column densities probed in the polarization spectrum map area
(between $\approx 10^{21.5}\,$cm$^{-2}$ and $\approx 10^{23.4}\,$cm$^{-2}$), 
we detect a change in \rdc/ from $\approx 1.2$ to $\approx 0.6$ (i.e., approximately 
a factor of two). In Figure \ref{f:pspecgraphs}c,
although the colored model curves do not go through the bulk of the data points, 
it can be seen that the change in \rdc/ expected from the models (approximately a factor of 1.5 to 2.0)
is similar to the change seen in the observations. 
Keeping in mind that the seven total intensity parameters
were set by considering only the column densities and temperatures derived from {\it Herschel} maps of the 
real cloud (see Section \ref{s:modcloud}), and that $R_t$, the sole
remaining parameter that can affect how much \rdc/ varies with column density, was also
set without any reference to the observed values of \rdc/ (see Section
\ref{s:modpol}), it seems surprising that our simple cloud model
can reproduce the general trends seen in the \rdc/ observations as well as it does.
We conclude that a simple cloud model which takes into account only ETAC 
can reasonably reproduce the observed systematic changes of the far-infrared 
polarization spectrum slope within the studied range of column densities.
This result is consistent with the RATs explanation for the observed changes in the 
polarization spectrum.  
It is important to note, however, that an increased degree of magnetic field disorder deep
inside the core 
provides an alternative way to obtain lower levels of polarization for light emitted 
from the densest central regions of a core.  We cannot discard this as an explanation 
for all or part of the trend we have discovered, 
even if the near-infrared spectro-polarimetry results discussed in Section \ref{s:introduction} 
suggest that in fact changes in grain alignment do play the dominant role.   
Disentangling field disorder from 
the expected loss of grain alignment due to RATs is a difficult problem that is perhaps 
best tackled with the help of MHD simulations \citep[e.g., see][]{2018king}.

In the context of the ETAC interpretation that is based on RATs theory,
the local temperature can be thought of as a proxy for the local intensity of the optical/near-IR radiation field and thus directly related to the dust grain alignment efficiency. 
Therefore, complementary to the above analysis of \rdc/ as a function of $N$,
it is also instructive to compare the observed relation of \rdc/ vs. temperature $T$
with the corresponding simple model curves (Figure \ref{f:pspecgraphs}d). The observed increase 
in \rdc/ with temperature is clearly well reproduced by the model curves.
Note that the temperature plotted in this figure is the LOS temperature rather than 
the local temperature (see Section \ref{s:compdata}).
We can associate each value of the transition radius $R_t$ with 
a corresponding LOS temperature $T_{m}(R_t)$. 
For instance, we find $T_{m}(R_t) = 31.3\,$K for $R_{t} = 0.6\,R$.
However, since it is the local temperature rather than the LOS temperature that 
serves as the better proxy for the local grain alignment efficiency, we can instead 
consider the local temperature $T_{l,m}(R_t)$ corresponding to the transition 
radius $R_{t}$. For $R_{t}= 0.3\,R$, $0.6\,R$, and $0.9\,R$, we find  $T_{l,m}(R_t)$
values of $21.4\,$K, $29.0\,$K, and $36.5\,$K, respectively. It would be interesting to test, by
applying a similar analysis to other cores of high column density, whether a critical local
temperature lying within this range is universal for high column density cores.

Note that we are not arguing here that high grain temperature is directly responsible for grain alignment.  Indeed, there is evidence against this.  Specifically, Globule 2 in the Southern Coalsack exhibits efficient grain alignment \citep{1984jones} despite its low temperature of $\approx10\,$K.  In the context of RATs theory, this may be attributed to the low column density of about $10^{22}\,$cm$^{-2}$, not enough to shield the cloud from the interstellar radiation field \citep{andersson2015}.  By way of comparison, we note that the peak column density in \roa/ is larger than $10^{23}\,$cm$^{-2}$.  Rather than arguing that grain temperature is directly related to grain alignment, we are instead suggesting that in \roa/ (and perhaps in other very dense cores) local temperature can serve as a proxy for the radiation intensity, and can thereby be related to grain alignment efficiency.    

The results presented in this paper introduce {\it a new method to probe the grain 
alignment efficiency in 
molecular clouds}, based on trends in the slope of the far-IR polarization spectrum. 
Provided a model is given for the studied cloud, one 
may test beyond which core depth, or below which local temperature, the grain alignment is 
no longer efficient. 
The usage of the polarization ratio as opposed to the polarization degree itself
(as has been done for various previous studies in the literature) offers an advantage, 
because the polarization degree is affected by inclination of the magnetic field 
lines with respect to the LOS, whereas the polarization ratio is not. 
This could explain why the trends in the observed relations 
of $p_C$ and $p_D$ as functions of $N$ (Figures 
\ref{f:pspecgraphs}a and \ref{f:pspecgraphs}b, respectively)
are more complicated than the trends observed in 
\rdc/ vs. $N$ (Figure \ref{f:pspecgraphs}c).
For instance, both $p_C$ and $p_D$ show 
a clear decrease as a function of $N$ in the range 
$22.0 < \log{N}\,(\mathrm{cm}^{-2})< 23.5$, 
However, for $\log{N}\,(\mathrm{cm}^{-2}) < 22.0$, there is
a wide spread in the values of $p_C$ and $p_D$, and the trend is no longer clear. This 
feature could potentially be
due to changes in the magnetic field inclination along the LOS.
The trends in \rdc/ are more clearly represented by a simple 
monotonic decrease as a function of $N$ (Figure \ref{f:pspecgraphs}c)
over the full range of column densities probed by these observations.
The polarization degree is also affected by unresolved field structure
in the plane-of-the sky, an effect that is also cancelled when using
the polarization ratio.
In the context of studying the role of the magnetic field
in star formation, developing new tools to probe grain alignment efficiency is critical for interpreting interstellar polarization
measurements arising from molecular cloud cores and filaments.

\subsection{Limitations of the simple spherical cloud model and comparison with longer wavelength polarimetry of \roa/}
\label{s:disc3}

In this initial work, we have used a very simple approach to model 
the cloud, but differences between the model and the real cloud are obvious when comparing 
column density and temperature maps (e.g., compare
Figures \ref{f:pspec}c and \ref{f:pspec}d to Figures \ref{f:modinputs}c 
and \ref{f:modinputs}d). 
There are numerous ways in which this model could be made more realistic,  
potentially improving the comparison between the model \rdc/ 
curves and the polarimetric observations. The most obvious improvement would
be to
abandon spherical symmetry, adopting a more complicated model
tailored to the real cloud.  Another way to add significant realism would be to introduce a gradual ``turn off" of the grain alignment efficiency as one 
moves deeper into the core, rather than using our strict 
cutoff at $r = R_{t}$. In addition, the number density profile (Plummer sphere) and 
the local temperature profile (linear) could be modified to match the 
{\it Herschel} data more accurately.
Finally, although the entire map shows far-infrared optical depth values less than unity, 
for the densest lines-of-sight this parameter reaches values high enough to invalidate
the
optically thin approximation used in Equation \ref{e:tflux}. 
For instance, at the peak column density of \roa/ we estimate optical depths of
0.53 and 0.21 for bands C and D, respectively. This could represent a 
difference of up to 20\% in \rdc/
at the peak column density, relative to the optically thin situation 
\citep[see, for example,][]{1989novak}.

Another extension of the work described here would be to expand the analysis to cover a wider wavelength range, including the \roa/ POL-2 data at $850\,\mu$m \citep[][]{kwon2018}.  This should be approached carefully as the HAWC\plus/ and POL-2
datasets likely probe very different column density regimes. In the context of the simple
spherical core model presented in this work, this can be verified by analyzing how the 
integrand of Equation \ref{e:tflux} (i.e., the dust emission per unit volume) varies as a function 
of the LOS depth $s$ toward the core center (i.e., for sightline $x = 0$). 
The HAWC\plus/ dust emission in bands
C and D peaks in the range $0.13 R - 0.26 R$ while the $850\,\mu$m dust emission peaks at a 
significantly deeper core layer, near $0.04 R$. This shows that POL-2 probably probes much closer
to the cold core center relative to HAWC\plus/ bands C and D. It seems unlikely that the simple sharp cutoff grain alignment prescription used here could capture the physical effects operating over such a large range of core depths.  
The investigation of the polarization spectrum over a wider range of wavelengths, probably using a more sophisticated grain alignment prescription,
is likely to prove informative, but such a study is beyond the scope of the present investigation. 

Finally, we note that after the present work was submitted for publication, we became aware of a
recent publication by \citet{2019pattle}, who used the same POL-2 data from \citet{kwon2018} to 
study the grain alignment efficiency in several dense regions within \ro/, including \roa/.  
Similarly to our work, their results highlight the importance of the incident radiation field for 
efficient alignment of dust grains.

\section{Summary and Conclusions}
\label{s:conclusions}

In this work we analyzed far-infrared polarimetric data from HAWC\plus//SOFIA in bands 
C ($89\,\mu$m) and D ($154\,\mu$m) for the densest portion of \ro/ (known as \roa/).
The main goal was to evaluate the changes in the slope of the polarization spectrum 
correlated with local cloud properties (more specifically, column densities and 
dust temperatures). From previous molecular cloud surveys of the far-infrared 
polarization spectrum, the slopes were typically found to be negative in this
same spectral range \citep{2002vai,2008vai,2012vai,2013zeng}. No systematic correlation 
between far-infrared polarization spectrum and cloud properties has been previously reported. 

We defined the polarization ratio \rdc/ = $p_{D}/p_{C}$ and investigated its distribution 
across \roa/. 
The polarization angles in bands C 
and D are very tightly correlated, which allowed the use of data for all sky positions 
for which measurements were available at both bands.
The polarization ratio map covers the surroundings of the massive star Oph S1, 
and also includes the peak density at the cloud core. 
We noticed a clear correlation of \rdc/ with $N$ and $T$:
in the range of column densities and temperatures covered by our dataset 
(approximately $2.8 \times 10^{21}\,\mathrm{cm}^{-2} < N < 2.5 \times 10^{23}\,\mathrm{cm}^{-2}$ 
and $20\,\mathrm{K} < T < 45\,\mathrm{K}$), \rdc/ decreases 
from $\approx 1.2$ (positive polarization spectrum 
slope) in the more diffuse portions of the core to approximately $0.6$ at 
the density peak (negative slope).
The discovery of positive polarization spectrum slopes is consistent with 
published dust grain 
models \citep{bethell2007,draine2009,2018guillet}. 

We explain the dependence of \rdc/ on $N$ and $T$ as a consequence of
the ETAC, i.e., grains in the warm and diffuse outskirts of the core are 
well aligned due to better exposure to radiation, while the 
alignment efficiency gradually decreases toward the colder and denser shielded core.
For the purpose of providing a sanity check on whether the ETAC can quantitatively
explain the magnitude of the observed change 
in \rdc/, we developed a very simple toy model for \roa/.
We model the cloud as a spherically symmetric core embedded in a uniform ambient medium,
and we determine the seven model parameters that determine the wavelength-dependent 
total intensity distributions
using {\it Herschel}-derived column density 
and temperature maps. We assume the simplest possible grain alignment efficiency
profile, i.e., the alignment is completely turned off interior to a certain radius 
from the core center. 
A range of values for cutoff radius is chosen based on the observed dependencies 
of $p_{D}$ and $p_{C}$ on $N$.  Finally, we compare the model's predictions for \rdc/
with the observed values, finding rough agreement.  
Based on this sanity check, we conclude that ETAC appears to be a plausible explanation for 
the polarization ratio trends observed.
We propose that the analysis of far-infrared polarization spectra 
can be used as a new method to probe the loss of grain alignment within dense 
interstellar cores.

\acknowledgments
The authors thank the anonymous referee for valuable suggestions that have improved the paper.
This work is based [in part] on observations made with the NASA/DLR Stratospheric Observatory for Infrared Astronomy (SOFIA). SOFIA is jointly operated by the Universities Space Research Association, Inc. (USRA), under NASA contract NAS2-97001, and the Deutsches SOFIA Institut (DSI) under DLR contract 50 OK 0901 to the University of Stuttgart.
We acknowledge the use of {\it Herschel} to conduct this work, including data from PACS. 
{\it Herschel} is an ESA space observatory with science instruments provided by European-led Principal Investigator consortia and with important participation from NASA.
JEV acknowledges support for this work provided by NASA through award \#SOF 05-0038 issued by USRA.
Portions of this work were carried out at the Jet Propulsion Laboratory, operated by the California Institute of Technology under a contract with NASA.
FPS acknowledges support by CAPES grant 2397/13-7.

\vspace{5mm}
\facilities{SOFIA} 

{\large\it Software:}
\texttt{python}, \texttt{Ipython} \citep{Perez2007}, \texttt{numpy} \citep{vanderWalt2011}, \texttt{scipy} \citep{Jones2001}, \texttt{matplotlib} \citep{Hunter2007}, \texttt{astropy} \citep{astropy:2013, astropy:2018}, 
LIC code (ported from publically-available IDL source by 
Diego Falceta-Gon\c{c}alves).

\bibliography{astroref}
\bibliographystyle{aa}

\end{document}